\documentclass[a4paper]{elsarticle}
\usepackage{graphicx}
\usepackage{setspace}
\usepackage{amsmath,amssymb}
\usepackage{color}
\usepackage{caption}
\usepackage{amsmath}
\usepackage{amssymb}
\usepackage{enumerate}
\usepackage{multirow}
\usepackage{array}
\usepackage{subfigure}
\usepackage{float}
\usepackage{geometry}
\usepackage{makecell}
\usepackage{lineno,hyperref}
\journal{Mathematics and Computers in Simulation}
\begin{document}

\title{A Coupled Two-relaxation-time Lattice Boltzmann-Volume Penalization method for Flows Past Obstacles}
\author[HEU]{Xiongwei Cui}

\author[HEU]{Zhikai Wang\corref{cor1}}
\cortext[cor1]{Corresponding author}
\ead{wangzhikai@hrbeu.edu.cn}

\author[HEU]{Xiongliang Yao}
\address[HEU]{College of Shipbuilding Engineering, Harbin Engineering University, Harbin,
 China 150001}

\author[ASU]{Minghao Liu}
\address[ASU]{School for Engineering of Matter, Transport and Energy, Arizona State University, Tempe, AZ 85281 USA}

\author[SDJU]{Fulin Yu}
\address[SDJU]{College of Naval Architecture and Port Engineering, Shandong Jiaotong University, Weihai, China 264200}

\begin{abstract}
    In this article, a coupled Two-relaxation-time Lattice Boltzmann-Volume penalization (TRT-LBM-VP) method is presented to simulate flows past obstacles. Two relaxation times are used in the collision operator, of which one is related to the fluid viscosity and the other one is related to the numerical stability and accuracy. The volume penalization method is introduced into the TRT-LBM by an external forcing term. In the procedure of the TRT-LBM-VP, the processes of interpolating velocities on the boundaries points and distributing the force density to the Eulerian points are unneeded. Performing the TRT-LBM-VP on a certain point, only the variables of this point are needed. As a consequence, the TRT-LBM-VP can be conducted parallelly. From the comparison between the result of the cylindrical Couette flow solved by the TRT-LBM-VP and that solved by the Single-relaxation-time LBM-VP (SRT-LBM-VP), the accuracy of the TRT-LBM-VP is higher than that of the SRT-LBM-VP. Flows past a single circular cylinder, a pair of cylinders in tandem and side-by-side arrangements, two counter-rotating cylinders and a NACA-0012 airfoil are chosen as numerical experiments to verify the present method further. Good agreements between the present results and those in the previous literatures are achieved.
\end{abstract}
\begin{keyword}
Two-relaxation-time model\sep Lattice Boltzmann\sep Immersed boundary method\sep Volume Penalization\sep Fluid-solid Interaction
\end{keyword}

\maketitle

\vspace{-6pt}

\section{\label{sec:intro}Introduction}
    As an alternative to the traditional Navier-Stokes (N-S) equation solver, the Lattice Boltzmann method (LBM) has been adopted widely to simulate flows past obstacles \cite{Shiyi1998}. The simplicity in coding, parallel computation and explicit calculation contribute to its popularity. In the procedure of simulating the flows past obstacles, the treatment of obstacle boundary is an extremely key point. Just like that in the traditional N-S equation solvers, the body-fitted grid method and the immersed boundary method (IBM) are the two main methods used in LBM when complex boundaries are involved.

    For the body-fitted grid method, generating a body-fitted grid is the first and important step. But this step is an extremely expensive process when complex boundaries are involved. Even with simple boundaries, creating a high quality body-fitted grid is of great difficulty. There is no doubt that both structured and unstructured grids are frequently used in the process of creating a body-fitted grid. However, the order of accuracy on the structured and unstructured grids is lower than that on the uniform Cartesian grids \cite{PIQUET2016272}. The IBM, proposed by Peskin \cite{peskin_2002}, can be implemented easily compared with the body-fitted grid method for there is no need to create a body-fitted grid. Two types of grids are used in the IBM: the Cartesian grid and the Lagrangian grid. The N-S equation is solved on the fixed Cartesian grid and the dynamics of the obstacles are represented by the Lagrangian grid. The information and variables on these two grids are related by a discrete delta function interpolation. Using a restoring force term to reflect the boundary effect on the flows is the extremely bright spot of the IBM. As a consequence, the N-S equations with a external forcing term can be solved on the whole Cartesian grid. Besides, there is also no need to regenerate a new body-fitted grid when the boundary of the obstacle moves or changes. Interpolating the velocity at boundaries and spreading the force density to the Eulerian points near the boundaries by using the delta function  are two main steps in the IBM. The IBM was incorporated into the LBM by Feng and Michaelides \cite{FENG2004602} to study the fluid-particles interaction problems. As mentioned above, the restoring force is used to reflect the boundary effect. But the non-slip boundary conditions are not always enforced. As a result, some streamlines may penetrate the fixed boundaries. To solve this problem, Wu and Shu \cite{Wu-Shu-I-V} proposed a implicit velocity correction-based IBM, in which the velocity corrections near the boundaries are considered as unknown and computed implicitly. From the results given by Wu and Shu, the non-slip boundary conditions are enforced very well. In the IBM, a Vandermonde matrix should be recomputed during every time step, meaning more computing time is needed \cite{PIQUET2016272}.

    Recently, Benamour et al. \cite{IJM294} incorporated another type IBM: Volume Penalization (VP), which is firstly proposed by Arquis and Caltagirone \cite{Arquis19841}, into LBM. In the VP, the obstacles are considered as a porous medium with extremely small permeability. The solid boundaries are modeled on the fixed Eulerian grid by using a mask function. Actually, the Lagrangian grids in the VP is part of the fixed Eulerian grid which are marked by the mask function. So one of the Lagrangian points coincides with a certain points of the fixed Eulerian grid. Compared with the direct forcing IBM and the velocity correction IBM, there is no need to interpolate the velocity at the boundaries and to spread the force density into the Eulerian points near the boundaries by using the delta function. Performing the VP procedure on a certain Lagrangian point just needs the variables on the Eulerain point with which the Lagrangian point coincides, meaning the whole VP procedure can be conducted parallelly. The VP has been successfully used by several researchers to solve the flows past obstacles \cite{ango1999, KEVLAHAN2001333, SCHNEIDER20051223}. Through the forcing term proposed by Guo \cite{PhysRevE.65.046308}, the VP is incorporated into LBM. Taking the parallelizability of the LBM into consideration, the whole LBM-VP procedure can be conducted parallelly.

    There are three typical collision operators in LBM: the Single-relaxation-time (SRT) collision operator, the Two-relaxation-time (TRT) collision operator and the Multiple-relaxation-time (MRT) collision operator. In terms of numerical accuracy and stability, these three operators differ from each other \cite{Luo-2011}. Maybe, the SRT LBM is the most popular LBM for its simplicity and easiness to implement \cite{Chen-Doolen-1998}. However, when complex geometries are involved, it may suffer from unphysical artifacts \cite{Le-Zhang}. Also, the accuracy in velocity distribution will fall when the single relaxation time $\tau$ increases \cite{Luo-2011}. To get a better accuracy and numerical stability, MRT collision operator was proposed \cite{MR1902782,Yan2014}. Compared with the SRT LBM, the numerical procedure of MRT LBM is more complicated. Besides, there is a challenging point in the MRT-LBM: selecting the multiple relaxation time parameters. Using two relaxation times for the even and odd order particle velocity moments, the TRT LBM not only maintains the simplicity of the SRT LBM but also haves the advantages of the MRT LBM in terms of accuracy and stability \cite{Luo-2011}. Coupled with the direct forcing and velocity correction IBM, the TRT LBM has been used to simulate flows past obstacles by Seta \cite{Seta-2014} and Hayashi et al. \cite{Hayashi-2012}. In this article, coupled with VP, the TRT LBM is used to simulate flows past obstacles.

    To validate the accuracy of the proposed TRT-LBM-VP, the cylindrical Couette flow is chosen as the experiment. Also the SRT-LBM-VP is used to solve the same experiment, the results of which are used to compare the accuracy of the TRT-LBM-VP and the SRT-LBM-VP. To verify the proposed TRT-LBM-VP further, flows past a single fixed circular, flows past a pair of cylinders in tandem and side-by-side arrangements, two counter-rotating cylinders and a NACA-0012 airfoil are conducted. The results are compared with the data in the previous literatures. The rest of this paper is arranged as follows. In Section \ref{sec:MNF}, the TRT LBM, Volume penalization and the incorporating of VP into TRT LBM are described. The whole computing procedure of the presented method is also given at the end of this section. The numerical experiments and the comparison of the results are shown in Section \ref{sec:ND}. In Section \ref{sec:Con}, some concluding remarks and recommendations for the future work are presented.

\section{\label{sec:MNF}Mathematical and Numerical Formulation}

In this section, the two-relaxation-times LBM is introduced firstly. Then the Volume Penalization method and the incorporation of VP method into the TRT-LBM are introduced in details. Finally, a whole computing procedure is given.

    \subsection{\label{sec:TRT-LBM}The two-relaxation-times LBM}
    Take the fluid-solid interaction (FSI) between incompressible viscous fluid and rigid obstacles into consideration. The dynamics of the fluid can be described by the Navier-Stokes equation (NS-equation):
    \begin{equation}
        {\frac{{\partial {\bf{u}}}}{{\partial t}}+ {\bf{u}}\cdot\nabla{\bf{u}}}+ {\frac{1}{\rho}}\nabla p = \mu \Delta {\bf{u}} + {\bf{f}}\label{eq:N-S}
    \end{equation}
    \begin{equation}
        \nabla\cdot {\bf{u}} = 0 ,\label{eq:continuity equation}
    \end{equation}
    where $\bf{u}$ is the velocity of the fluid, $\mu$ is the dynamic viscosity, $\rho$ is the density, $p$ is the pressure and ${\bf{f}}$ is the body force. The above two equations are the macroscopic mass  and momentum equations of incompressible viscous fluids. In the Lattice Boltzmann framework, the motion of the fluids is governed by the following kinetic equations, without the body force term, for the distribution function $f_k$ \cite{Luolishi1997}:
    \begin{equation}
        \frac{{\partial {f_\alpha }}}{{\partial t}} + {\bf{e}^\alpha } \cdot \nabla {f_\alpha } = {\Omega _\alpha }\label{eq:LBM}
    \end{equation}
    in which $f_\alpha$ is the particle distribution function; $t$ is the time; ${{\bf{e}}_\alpha }$ is the particle velocity in the $\alpha {\rm{th}}$ direction; ${\Omega _\alpha}$ is the collision operator. For the nine-velocity lattice model in two dimension (D2Q9), the discrete velocity vectors can be defined as:
    \begin{equation}
        {{\bf{e}}_\alpha } = \left\{ {\begin{array}{*{20}{c}}
        {(0,0),}\\
        {(cos{\theta _\alpha },sin{\theta _a}),}\\
        {\sqrt 2 (cos{\theta _\alpha },sin{\theta _a}),}
        \end{array}} \right.\begin{array}{*{20}{c}}
        {\begin{array}{*{20}{c}}
        {\alpha  = 0}\\
        {{\theta _a} = {{\left( {\alpha  - 1} \right)\pi } \mathord{\left/
         {\vphantom {{\left( {\alpha  - 1} \right)\pi } 2}} \right.
         \kern-\nulldelimiterspace} 2},\alpha  = 1 - 4}\\
        {{\theta _a} = {{\left( {2\alpha  - 9} \right)\pi } \mathord{\left/
         {\vphantom {{\left( {2\alpha  - 9} \right)\pi } 4}} \right.
         \kern-\nulldelimiterspace} 4},\alpha  = 5 - 8}
        \end{array}}.
        \end{array}\label{eq:D2Q9}
    \end{equation}
    In the TRT-LBM model, two relaxation times are used in the collision operator, which is written as\cite{GINZBURG20051171}:
    \begin{equation}
        {\Omega _\alpha } =  - \frac{{f_\alpha ^s - f_\alpha ^{eq\_s}}}{{{\tau _s}}} - \frac{{f_\alpha ^a - f_\alpha ^{eq\_a}}}{{{\tau _a}}},
        \label{eq:TRT-collision}
    \end{equation}
    where superscripts $s$ and $a$ represent the symmetrical and antisymmetrical parts of the particle distribution functions. And the symmetrical and antisymmetrical particle distribution functions are defined as:
    \begin{equation}
        f_\alpha ^s = \frac{{{f_\alpha } + {f_{ - \alpha }}}}{2},\quad f_\alpha ^{a}    = \frac{{{f_\alpha } - {f_{ - \alpha }}}}{2},
    \end{equation}
    and their equilibrium distribution functions are:
    \begin{equation}
        f_\alpha ^{eq\_s} = \frac{{f_\alpha ^{eq} + f_{ - \alpha }^{eq}}}{2}, \quad f_\alpha ^{eq\_a} = \frac{{f_\alpha ^{eq} - f_{ - \alpha }^{eq}}}{2},
    \end{equation}
    in which the subscript $- \alpha$ is the direction opposite to $\alpha$, for instance, ${f_{ - 1}} = {f_3}$ and $f_{ - 3}^{eq} = f_1^{eq}$ in the D2Q9 model. Specially, it should be noted that ${f_{ - 0}} = {f_0}$ and $f_{ - 0}^{eq} = f_0^{eq}$. In the D2Q9 model, the equilibrium distribution functions are defined as:
    \begin{equation}
        f_\alpha ^{eq} = \rho {w_\alpha }\left[ {1 + \frac{{{{\bf{e}}_\alpha } \cdot {\bf{u}}}}{{c_s^2}} + \frac{{{{\left( {{{\bf{e}}_a} \cdot {\bf{u}}} \right)}^2} - c_s^2{{\left| {\bf{u}} \right|}^2}}}{{2c_s^4}}} \right],\label{eq:feq}
    \end{equation}
    in which ${c_s} = {1\mathord{\left/{\vphantom {1{\sqrt3}}}\right.\kern-\nulldelimiterspace}{\sqrt 3}}$ is the sonic speed and the weight factors are ${w_0} = {4 \mathord{\left/{\vphantom {4 9}} \right.\kern-\nulldelimiterspace} 9}$, ${w_{1 - 4}} = {1 \mathord{\left/{\vphantom {1 9}} \right.\kern-\nulldelimiterspace} 9}$ and ${w_{5 - 8}} = {1 \mathord{\left/{\vphantom {1 {36}}}\right.\kern-\nulldelimiterspace} {36}}$. ${\tau _s}$ and ${\tau _a}$ are relaxation times. The relation between ${\tau _s}$ and the kinematic viscosity $\upsilon $: ${\tau _s}  = {\upsilon  \mathord{\left/{\vphantom {\upsilon  {\left( {c_s^2\Delta t} \right)}}} \right.\kern-\nulldelimiterspace} {\left( {c_s^2\Delta t}\right)}}$ can be used to get the value of ${\tau _s}$. Through a magic parameter\cite{DHUMIERES2009823}, the relaxation parameter ${\tau _a}$ can be obtained from the following equation:
    \begin{equation}
        \Lambda  = \left( {{\tau _s} - \frac{1}{2}} \right)\left( {{\tau _a} - \frac{1}{2}} \right).
    \end{equation}
    The magic parameter influences the stability of the TRT-LBM scheme\cite{DHUMIERES2009823}. In order to maintain the most stability, the magic parameter is set as $1/4$\cite{Hayashi-2012}. The macroscopic density, $\rho $, and velocity, ${\bf{u}}$ are defined as follows:
    \begin{equation}
        \rho  = \sum\limits_\alpha  {{f_\alpha }} \label{eq:rho}
    \end{equation}
    \begin{equation}
        \rho {\bf{u}} = \sum\limits_\alpha  {{{\bf{e}}_\alpha }{f_a}} . \label{eq:rho*velocity}
    \end{equation}

    The Eq. (\ref{eq:LBM}) can be split into two sub-steps \cite{LeeT2003JCP}: collision
    \begin{equation}
        {g_\alpha } = {f_\alpha } + {\Omega _\alpha }\label{eq:collision}
    \end{equation}
    and streaming
    \begin{equation}
        {f_\alpha }\left( {{\bf{x}},t + \Delta t} \right) = {g_\alpha }\left( {{\bf{x}} - \Delta {{\bf{x}}_\alpha },t} \right).\label{eq:streaming}
    \end{equation}

    The effect of the external force can be reflected by adding an external forcing term to the collision sub-step. Similarly, the forcing term is also divided into symmetrical part $F_\alpha ^s$ and antisymmetrical part $F_\alpha ^a$, which are defined as\cite{Seta.PhysRevE89}:
    \begin{equation}
        F_\alpha ^s = \frac{{{F_\alpha } + {F_{ - \alpha }}}}{2}, \quad F_\alpha ^a = \frac{{{F_\alpha } - {F_{ - \alpha }}}}{2}.
    \end{equation}
    Here, the external forcing term proposed by Guo et al.\cite{Guo2002-LBM-force} is adopted:
    \begin{equation}
        {F_\alpha } = \left( {1 - \frac{1}{{2\lambda }}} \right){w_\alpha }\left( {\frac{{{{\bf{e}}_\alpha } - {\bf{u}}}}{{c_s^2}} + \frac{{{{\bf{e}}_\alpha } \cdot {\bf{u}}}}{{c_s^4}} \cdot {{\bf{e}}_\alpha }} \right) \cdot {{\bf{f}}}, \label{eq:external-force-term}
    \end{equation}
    \begin{equation}
        \rho {\bf{u}} = \sum\limits_\alpha  {{{\bf{e}}_\alpha }{f_\alpha }}  + \frac{1}{2}{{\bf{f}}} \cdot \Delta t. \label{eq:rho*velocity-ex}
    \end{equation}
    The collision sub-step is modified as:
    \begin{equation}
        {g_\alpha } = {f_\alpha } + {\Omega _\alpha } + \Delta t\left( {1 - \frac{1}{{2{\tau _s}}}} \right)F_\alpha ^s + \Delta t\left( {1 - \frac{1}{{2{\tau _a}}}} \right)F_\alpha ^a. \label{eq:collision-modified}
    \end{equation}

    \subsection{\label{sec:VP} The Volume Penalization Method and the Incorporation of VP into LBM}

\begin{figure}[h]
        \centering
        \includegraphics[width=6.5cm]{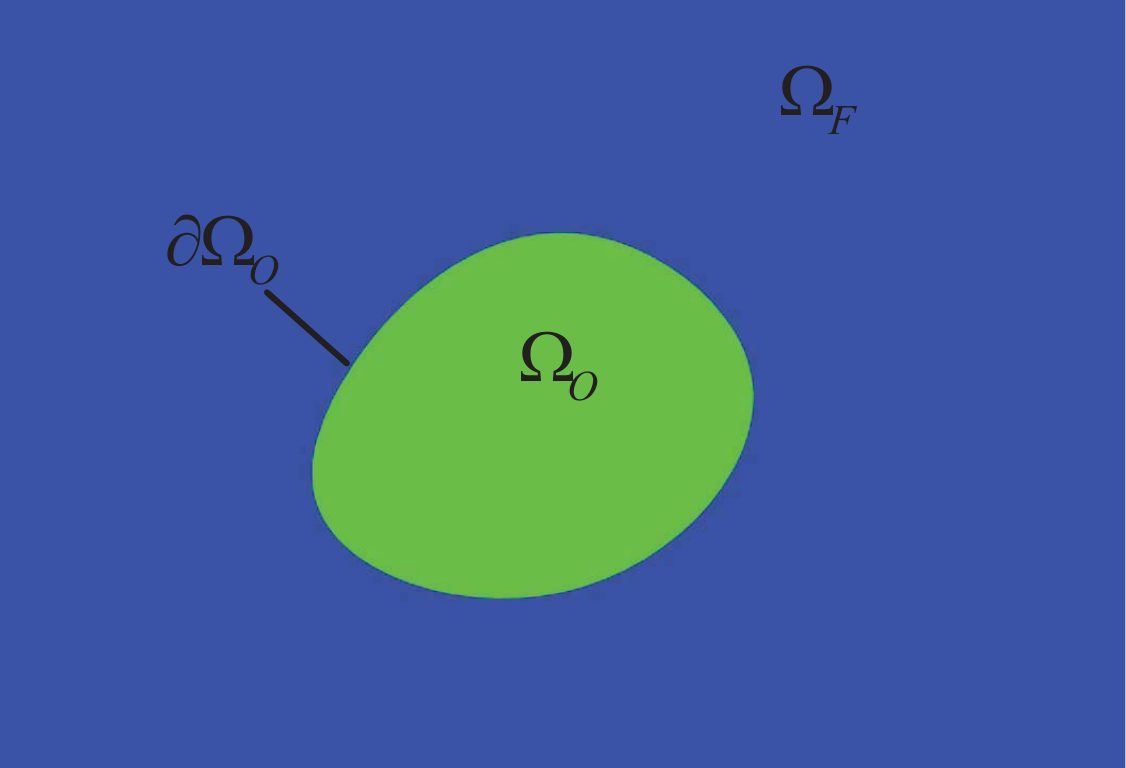}
        \caption{The computational domain of obstacle and fluid}\label{fig:C-O-F}
    \end{figure}

    Let us consider an obstacle immersed in the fluid, as shown in the Fig. \ref{fig:C-O-F}. On the boundary of the obstacle domain ${\Omega _O}$, the non-slip boundary condition can be described as:
    \begin{equation}
        {\bf{u}}\left| {_{\partial {\Omega _O}}} \right. = {{\bf{U}}_O},\label{eq:boundary-condition}
    \end{equation}
    where $\partial {\Omega _O}$ is the boundary of the obstacles and ${\bf{U}}_O$ is the velocity of the obstacles. For problems involving only fixed obstacles are considered in the present method, the velocity of the obstacles ${\bf{U}}_O$ is equal to zero. $\Omega _F$ is the fluid domain. The union of these two domains $\Omega  = {\Omega _F} \cup {\Omega _O}$ is the entire domain.

    The Dirichlet problem Eq. (\ref{eq:N-S}, \ref{eq:continuity equation}, \ref{eq:boundary-condition}) can be solved by the volume penalization method \cite{Angot1999,Carbou2003Boundary}. In the volume penalization method, the solid obstacles are modeled as porous media. By adding a penalization term on the velocity, the momentum Eq. (\ref{eq:N-S}) is modified as:
    \begin{equation}
        \frac{{\partial {\bf{u}}}}{{\partial t}} + {\bf{u}} \cdot \nabla {\bf{u}} =  - \frac{1}{\rho }\nabla p + \mu {\nabla ^2}{\bf{u}} + {{\bf{F}}_b} - \frac{{{\bf{\chi }}\left( {{\bf{x}},t} \right)}{\rho }}{\eta }\left( {{\bf{u}} - {{\bf{U}}_O}} \right),\label{eq:N-S-modified}
    \end{equation}
    where
    \begin{equation}
        \chi \left( {{\bf{x}},t} \right) = \left\{ {\begin{array}{*{20}{c}}
        1\\
        0
        \end{array}} \right.\begin{array}{*{20}{c}}
        {}&{\begin{array}{*{20}{c}}
        {\bf{x} \in {\Omega _O}}\\
        {other}
        \end{array}}
        \end{array}\label{eq:mask function}
    \end{equation}
    is the mask function used to describe the obstacle geometry and $\eta  \ll 1$ is the penalization parameter. It can be seen that there is no Dirichlet boundary condition in Eq. (\ref{eq:N-S-modified}). The solution of the penalized N-S Eq. (\ref{eq:N-S-modified}) tends towards the exact solution of N-S equation imposing no-slip boundary conditions with $\eta  \to 0$ \cite{Angot1999,Carbou2003Boundary,Engels201596}. The hydrodynamic forces acting on the fixed obstacle can be obtained through:
    \begin{equation}
        {{\bf{F}}_O} = \mathop {\lim }\limits_{\eta  \to 0} \int\limits_\Omega  {\frac{\chi }{\eta }\left( {{\bf{u}} - {{\bf{U}}_O}} \right)d\Omega }  = \mathop {\lim }\limits_{\eta  \to 0} \int\limits_{{\Omega _O}} {\frac{\chi {\rho }}{\eta }\left( {{\bf{u}} - {{\bf{U}}_O}} \right)d\Omega }.\label{eq:hydrodynamic force}
    \end{equation}
    Practically, the parameter $\eta$ can be set as a very small value, for instance, $\eta  = {10^{ - 8}}$. So the effect of the obstacle on the fluid can be reflected by the force density:
    \begin{equation}
        {{\bf{F}}_{VP}} =-\frac{{\chi \rho }}{\eta }\left( {{\bf{u}} - {{\bf{U}}_O}} \right).\label{eq:VP-force}
    \end{equation}
    From Eq. (\ref{eq:rho*velocity-ex}), the fluid velocity  consists of two parts \cite{Wu-Shu-I-V}. The density distribution functions contribute to the first part, which is represented by the intermediate velocity ${{\bf{u}}^ * }$. And the external force term contributes to the second part, which can be taken as a velocity correction $\delta {\bf{u}}$. The intermediate velocity can be expressed as:
    \begin{equation}
        \rho {{\bf{u}}^*} = \sum\limits_\alpha  {{{\bf{e}}_\alpha }{f_\alpha }}\label{eq:rho*velocity-star}
    \end{equation}
    and the velocity correction as
    \begin{equation}
        \rho \delta {\bf{u}} = \frac{1}{2}{{\bf{F}}_{VP}}\Delta t. \label{eq:rho*delta-velocity}
    \end{equation}
    And Eq. (\ref{eq:rho*velocity-ex}) is rewritten as:
    \begin{equation}
        {\bf{u}} = {{\bf{u}}^*} + \delta {\bf{u}}. \label{eq:velocity-r}
    \end{equation}

    Substituting Eq. (\ref{eq:velocity-r}) into Eq. (\ref{eq:VP-force}), the force density, contributed from the immersed boundary, can be obtained:
    \begin{equation}
        {{\bf{F}}_{VP}} = -\frac{{\chi \rho }}{\eta }\left( {{{\bf{u}}^ * } + \delta {\bf{u}} - {{\bf{U}}_O}} \right). \label{eq:VP-force-1}
    \end{equation}
    Then substituting Eq, (\ref{eq:VP-force-1}) into Eq. (\ref{eq:rho*delta-velocity}), the velocity correction can be calculated through \cite{IJM294}:
    \begin{equation}
        \delta {\bf{u}} = \frac{{\chi \left( {{{\bf{U}}_O} - {{\bf{u}}^ * }} \right)}}{{\frac{{2\eta }}{{\Delta t}} + \chi }}
        \label{eq:d-V}
    \end{equation}
    and the force term reflecting the effect of the immersed boundary can be expressed as:
    \begin{equation}
        {{\bf{F}}_{VP}} = \frac{{2\rho \chi \left( {{{\bf{U}}_O} - {{\bf{u}}^ * }} \right)}}{{2\eta  + \chi \Delta t}}. \label{eq:force-density}
    \end{equation}

    In summary, the whole steps of the algorithm are as follows.
    \begin{enumerate}[\indent(1)]
        \item Design the computational grid, and arrange initial values on the computational grid.
        \item Use Eq. (\ref{eq:collision-modified}) to obtain the density distribution function after the collision sub-step of the $t = {t_n}$ step (initially setting $F_\alpha ^s = 0$ and $F_\alpha ^a = 0$). \label{step:2.2}
        \item Performing the stream sub-step using Eq. (\ref{eq:streaming})
        \item Use Eq. (\ref{eq:rho}) and Eq. (\ref{eq:rho*velocity-star}) to obtain the macroscopic density and intermediate velocity.
        \item Use Eq. (\ref{eq:d-V}) and Eq. (\ref{eq:force-density}) to obtain the velocity corrections and the force density.
        \item Correct the fluid velocity on using Eq. (\ref{eq:velocity-r}) and compute the two relaxation-time collision operator using Eq. (\ref{eq:TRT-collision}). \label{step:2.5}
        \item Repeat step (\ref{step:2.2}) to step (\ref{step:2.5}) until the convergence is reached.
    \end{enumerate}
\section{\label{sec:ND}Numerical Results and Discussions}

    In this article, some numerical experiments are conducted to verify the present TRT-LBM-VP method. Firstly, the cylindrical Couette flow experiments solved by the TRT-LBM-VP and the SRT-LBM-VP are conducted. The results are used to compare the accuracy of the two methods. Then some other experiments involving incompressible viscous flows past a single fixed circular cylinder and flows past a pair of circular cylinders in tandem and side-by-side arrangement are carried out the validate the present method further. To validate the present method for some more complex boundaries, flows past a pair of counter-rotating cylinders with different angular velocities and flows past NACA-0012 airfoil with $0^\circ$ angle of attack (AOA) and $10^\circ$ AOA are carried out then. The results from previous literatures serve as references for the comparison.

    In the following numerical experiments, the Reynolds number (Re) is defined as:
    \begin{equation}
        {\mathop{\rm Re}\nolimits}  = \frac{{{U_\infty }D}}{\upsilon },
    \end{equation}
    in which ${U_\infty }$ is the free stream velocity and $D$ is the diameter of the cylinder. $\upsilon $ is the kinematic viscosity of the fluid. The drag coefficient ${C_d}$ and lift coefficient ${C_l}$ of cylinder are defined as
    \begin{equation}
        {C_d} = \frac{{2{F_d}}}{{\rho U_\infty ^2D}} \quad {C_l} = \frac{{2{F_l}}}{{\rho U_\infty ^2D}},
    \end{equation}
    where ${F_d}$ and ${F_l}$ are the drag force and lift force respectively. In the VP, the drag force and lift force can be calculate through:
    \begin{equation}
        {F_d} = \sum {F_{VP}^x\Delta x\Delta y}\quad  {F_l} = \sum {F_{VP}^y\Delta x\Delta y},
    \end{equation}
    where $F_{VP}^x$ and $F_{VP}^y$ are the $x$ and $y$ component of the force density in Eq. (\ref{eq:force-density}) respectively.
    For the unsteady cases where the vortex shedding happens, the Strouhal number is defined as:
    \begin{equation}
        St = \frac{{{f_q}D}}{{{U_\infty }}},
    \end{equation}
    where ${f_q}$ is the vortex shedding frequency.

    \begin{figure}[H]
        \centering
        \includegraphics[width=8cm]{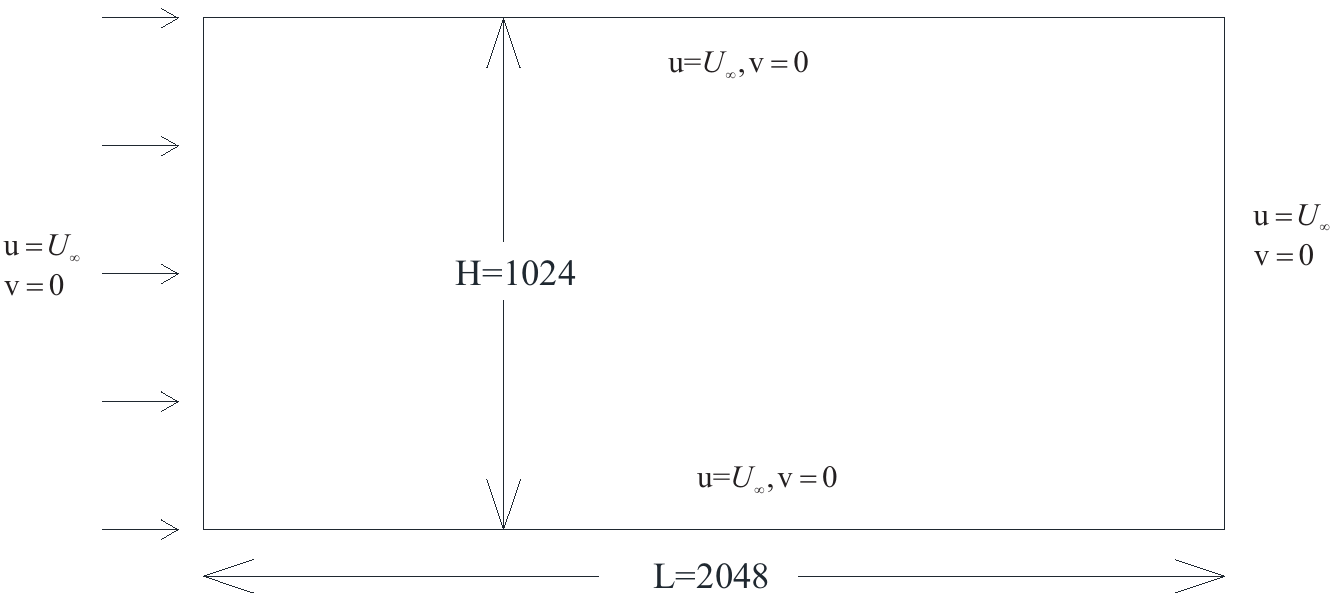}
        \caption{The computational domain and boundary conditions for flow past obstacles (except the cylindrical Couette flow experiments)}\label{fig:CD}
    \end{figure}

    For the all flow past obstacles numerical experiments, the density $\rho$ of the fluid is set as $1.0$. In the flows past one and two circular cylinders experiments, the diameter of cylinder $D$ is $40$. The computational domain of all numerical experiments is a rectangle domain with the height $H$ of $1024$ and the length $L$ of 2048, as shown in the Fig. \ref{fig:CD}
    \subsection{\label{sec:cylindrical-couette} The cylindrical Couette flow}
\begin{figure}[H]
        \centering
        \includegraphics[width=8cm]{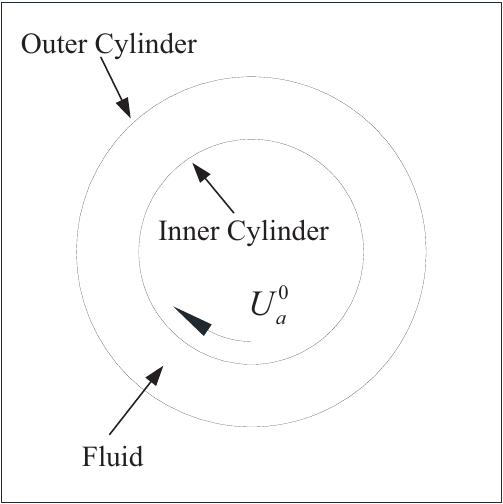}
        \caption{The computational domain and boundary conditions for the cylindrical Couette flow experiments}\label{fig:CD-c-c}
    \end{figure}

    For this case, two concentric cylinders are enclosed in a square computational domain, as shown in Fig. \ref{fig:CD-c-c}. The length of the square domain $L$ is set as 1. The radii of the outer cylinder $R_I$ and the inner cylinder $R_O$ are set as $70/200$ and $45/200$ respectively. The centers of the two cylinders are located at the center of the square domain. The outer cylinder is at rest, and the tangential velocity of the inner cylinder $U_a^0$ is set as $0.001$. The Reynolds number is 10.

\begin{figure}[H]
        \subfigure[Azimuthal velocity profile along the horizontal place through the center of the cylinders]{
            \label{fig:u-a-cylindrical-couette}
            \includegraphics[width=8.5cm]{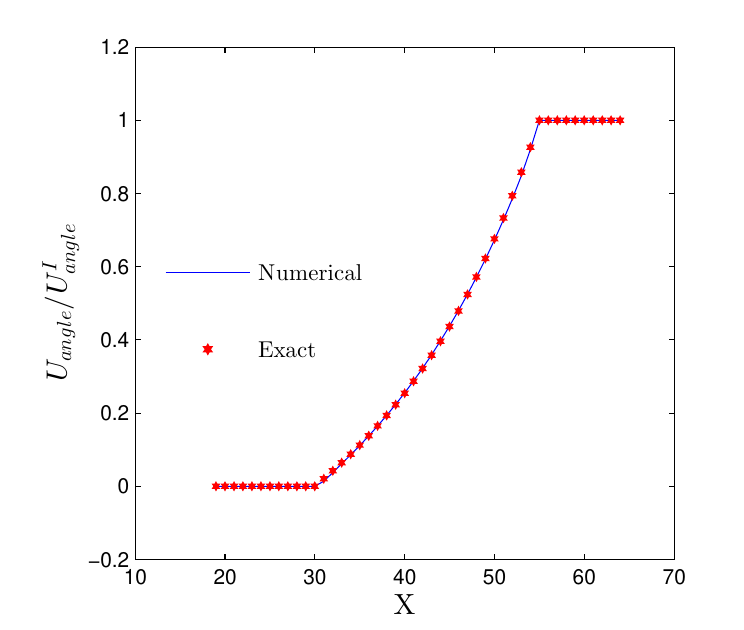}}
        \subfigure[Velocity contour]{
            \label{fig:v-contour-cylindrical-couette}
            \includegraphics[width=7cm]{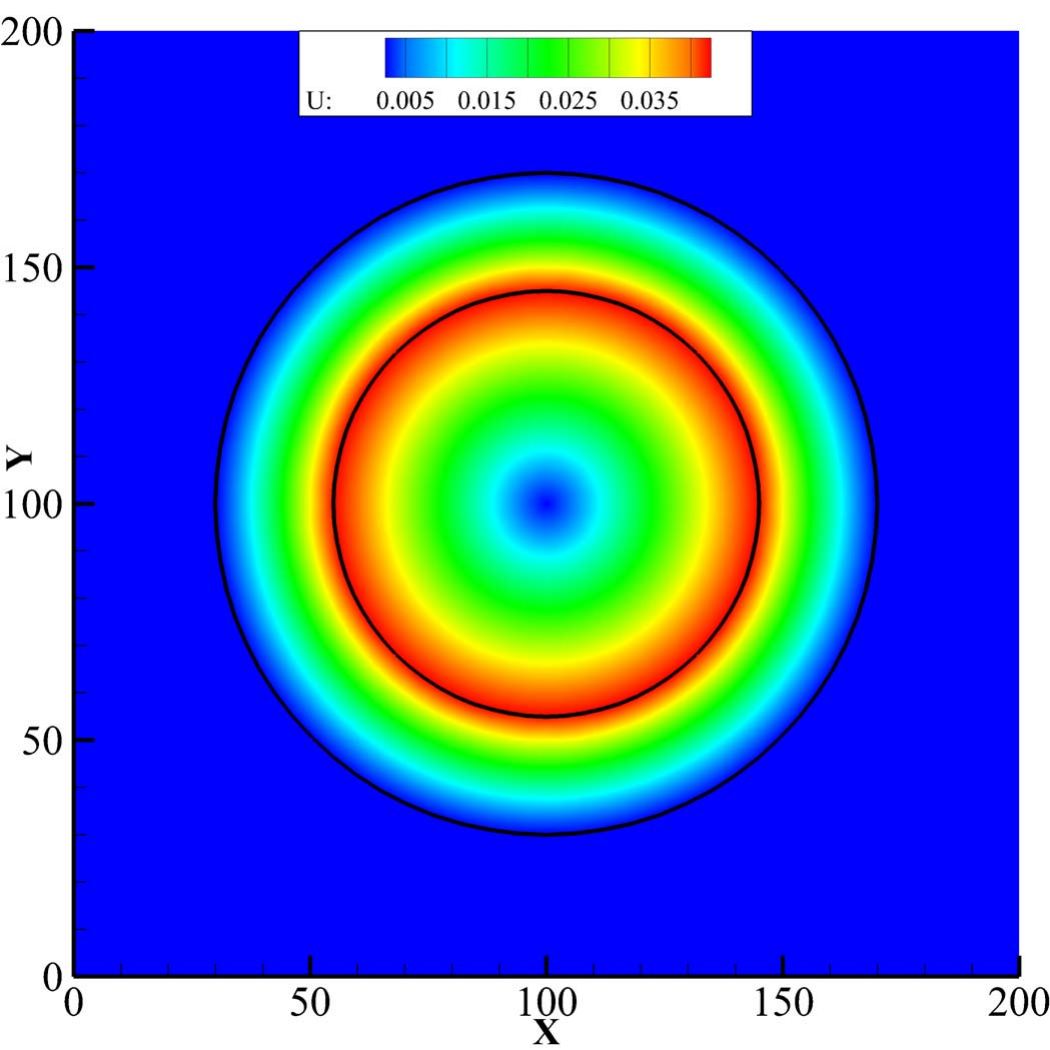}}
        \caption{Velocity contour and azimuthal velocity profile between two concentric cylinders using $200\times200$ grid sized.}
        \label{fig:couette-flow}
    \end{figure}

    The flow between the two cylinders will reach a steady state. And the analytical solution of the azimuthal velocity $U_a$ is given by \cite{Taylor-cylindrical-1923}
    \begin{equation}
        {U_a} = \frac{{\frac{{R_I^2R_O^2}}{{{r^2}}} - R_I^2}}{{R_O^2 - R_I^2}}U_a^0,
    \end{equation}
    where $r$ is the radial coordinate. In Fig. \ref{fig:couette-flow}, the velocity contour and azimuthal velocity profile along the horizontal plane through the center of the cylinders are given. From the azimuthal velocity profile, a good agreement can be obtained between the numerical result and the analytical solution.

\begin{figure}[H]
        \centering
        \includegraphics[width=8cm]{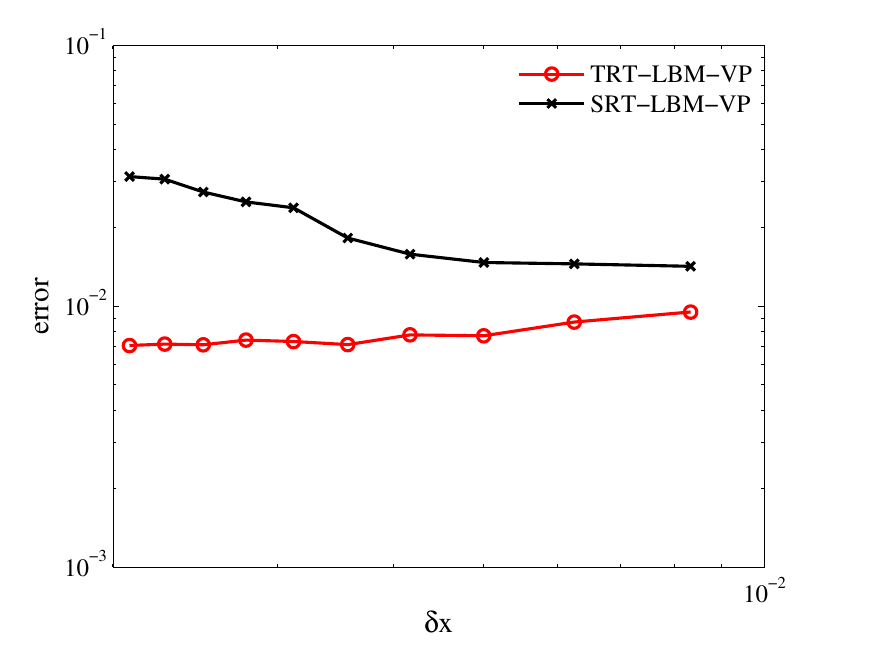}
        \caption{Error in azimuthal velocity versus for $\delta x = 1/120$, $1/160$, ..., $1/480$.}\label{fig:error-cylindrical-couette}
    \end{figure}

    Also the SRT-LBM-VP is used to solve the same problem. In Fig. \ref{fig:error-cylindrical-couette}, the errors in azimuthal velocity predicted using the proposed TRT-LBM-VP and the SRT-LBM-VP are given. It is obvious that the errors in the proposed TRT-LBM-VP are smaller than that in the SRT-LBM-VP, which means the accuracy of the TRT-LBM-VP is higher than that of the STR-LBM-VP.

    \subsection{\label{sec:sigle-cylinder} Flows past a fixed circular cylinder}

\begin{figure}[H]
            \subfigure[Streamlines and velocity contours for $Re=20$]{
                \label{fig:stream-re=20}
                \includegraphics[width=7cm]{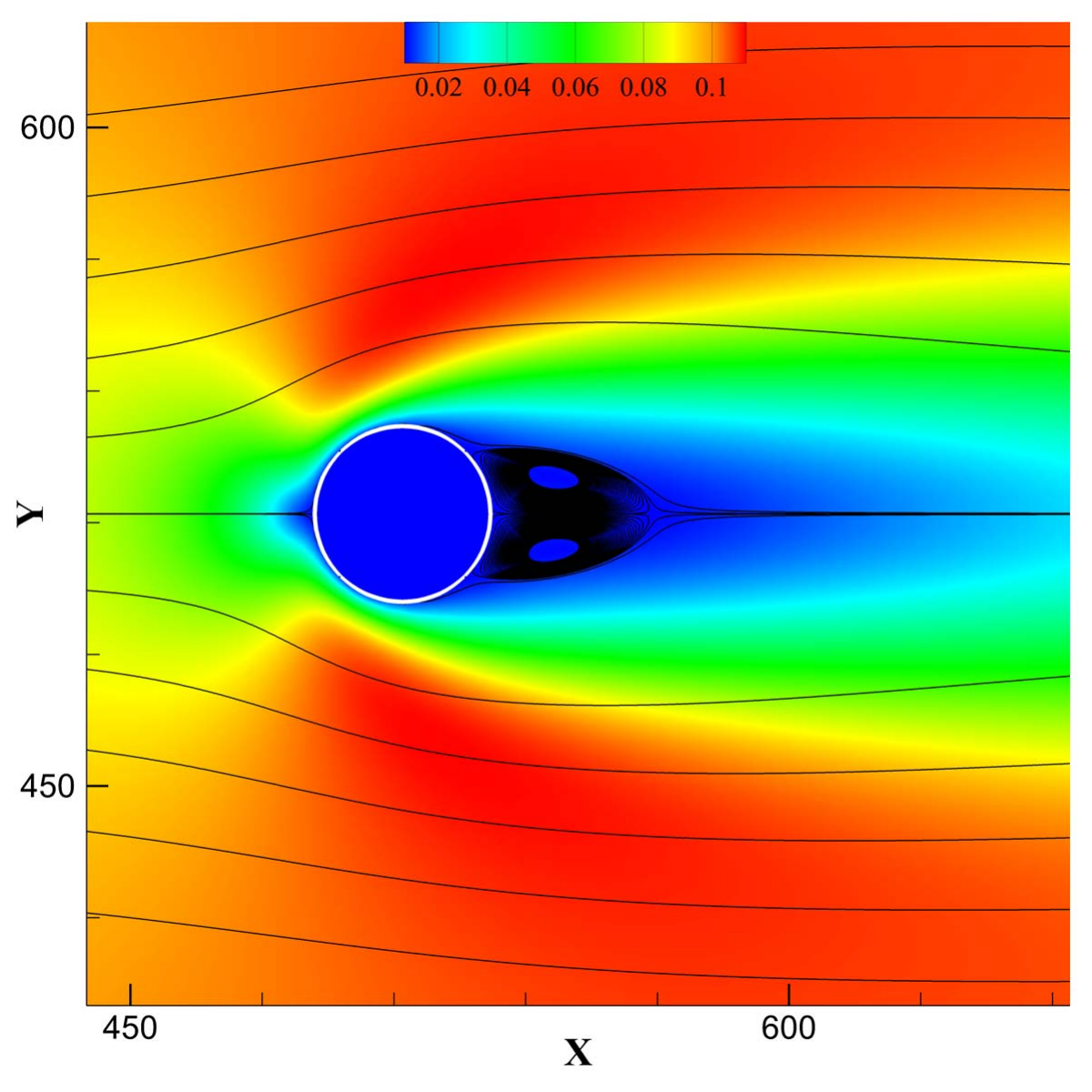}}
            \subfigure[Streamlines and velocity contours for $Re=40$]{
                \label{fig:stream-re=40}
                \includegraphics[width=7cm]{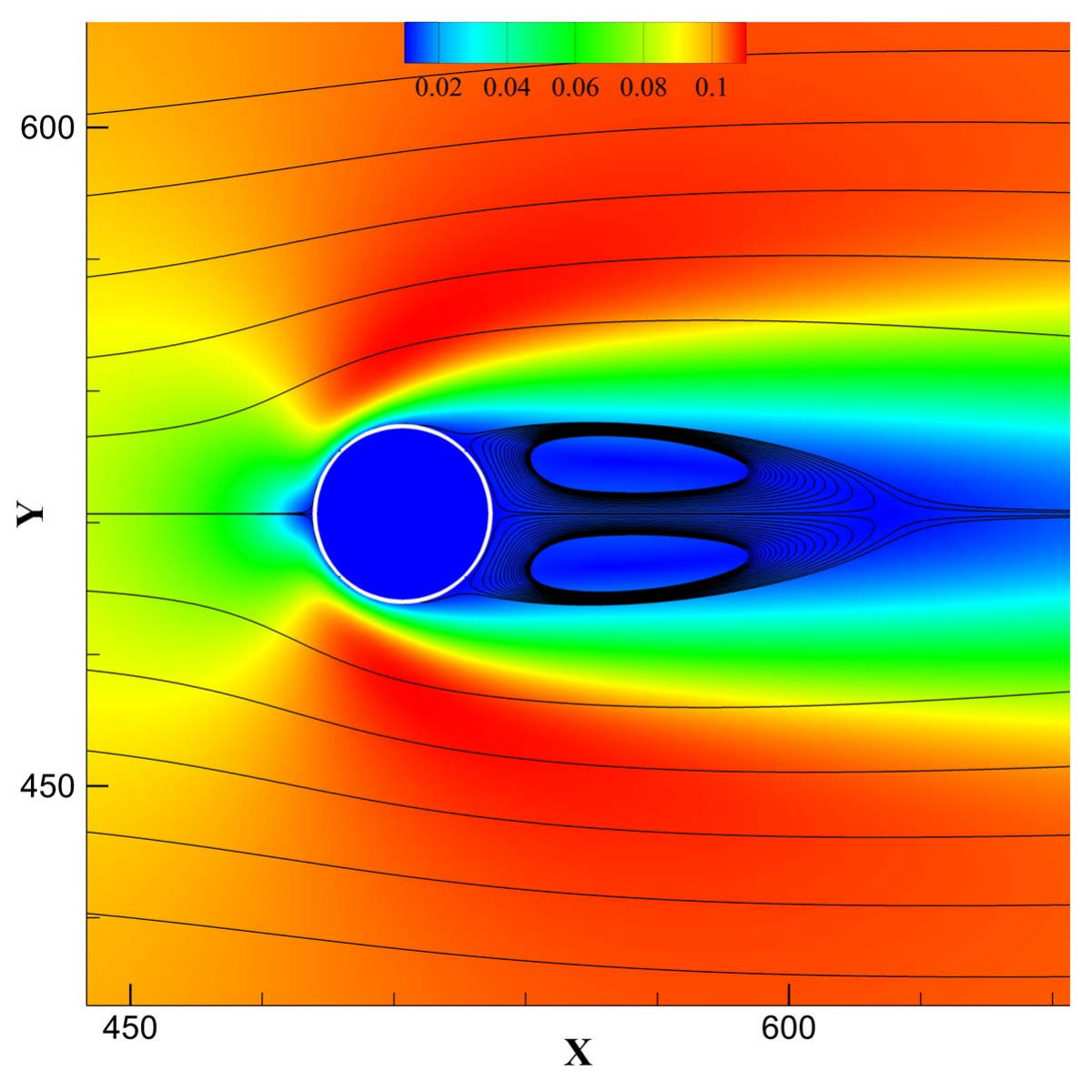}}
            \caption{Steady flows past a cylinder at ${\mathop{\rm Re}\nolimits}=20$ and ${\mathop{\rm Re}\nolimits}=40$.}
            \label{fig:figure.low-Re}
        \end{figure}

        The center of the fixed circular cylinder is located at $(H/2,H/2)$. For the cases of $Re=20$ and $Re=40$, the flows around the cylinder reach a steady state. A pair of symmetrical stationary recirculating eddies develop behind the cylinder, as shown in Fig. \ref{fig:figure.low-Re}. The distance from the end of the wake to the nearmost point of the cylinder will increase when the Reynolds number grows. The drag coefficient $C_d$, length of the recirculation region $L$ (scaled by $D$) and the separation angle $\theta _s$ are compared with the data in previous literatures in Table \ref{tab:low-Re}. There is a good agreement between the present results and those in previous literatures.

        \begin{table}[H]
            \centering
                \captionsetup{justification=centering}
            \caption{Drag coefficient, length of bubbles and separation angle for flows past a cylinder at ${\mathop{\rm Re}\nolimits}  = 20$ and ${\mathop{\rm Re}\nolimits}  = 40$.}
            \begin{tabular}{p{1.7cm}<{\centering}|p{4.4cm}<{\centering}|p{2.5cm}<{\centering}|p{2.5cm}<{\centering}|p{2.5cm}<{\centering}}
              \hline
              Case & References & ${C_d}$ & ${L_s}$ & ${\theta _s}$ \\
              \hline \multirow{4}{*}{${\mathop{\rm Re}\nolimits}  = 20$} & Niu et al. \cite{Niu-2006} & $2.144$ & $0.945$ & $42.9^{\circ}$\\
              \cline{2-5} & J. Wu and C. Shu \cite{Wu-Shu-I-V} & $2.091$ & $0.93$ & - \\
              \cline{2-5} & X. Cui et al. \cite{CUI201724} & $2.11$ & $0.93$ & $42.8^{\circ}$ \\
              \cline{2-5} & Present & $2.09$ & $0.93$ & $43.0^{\circ}$ \\
              \hline \multirow{4}{*}{${\mathop{\rm Re}\nolimits}  = 40$} & Niu et al. \cite{Niu-2006} & $1.589$ & $2.26$ & $53.86^{\circ}$ \\
              \cline{2-5} & J. Wu and C. Shu \cite{Wu-Shu-I-V} & $1.539$ & $2.23$ & - \\
              \cline{2-5} & X. Cui et al. \cite{CUI201724} & $1.56$ & $2.20$ & $53^{\circ}$ \\
              \cline{2-5} & Present & $1.55$ & $2.25$ & $52.9^{\circ}$ \\
              \hline
            \end{tabular}
            \label{tab:low-Re}
        \end{table}

\begin{figure}[H]
            \centering
            \subfigure[${\mathop{\rm Re}\nolimits}=100$]{
                \label{fig:figure.5.a}
                \includegraphics[width=7cm]{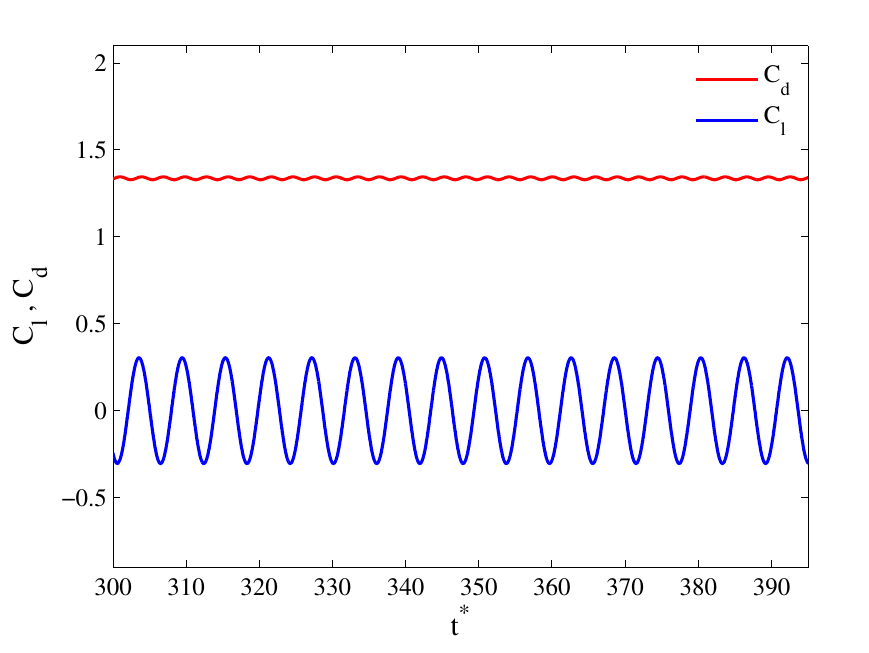}}
            \subfigure[${\mathop{\rm Re}\nolimits}=200$]{
                \label{fig:figure.5.b}
                \includegraphics[width=7cm]{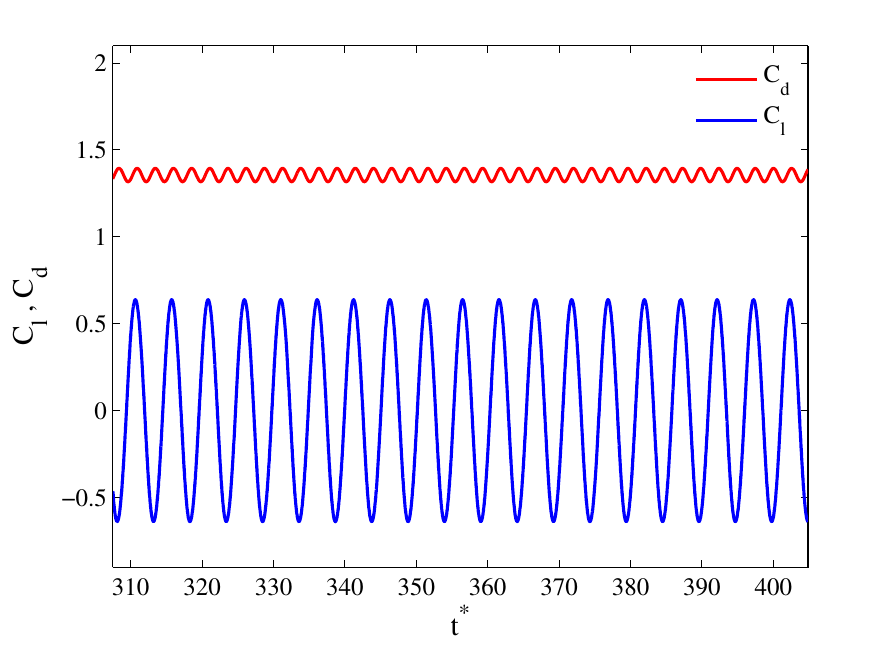}}
            \caption{Evolution of drag and lift coefficients at ${\mathop{\rm Re}\nolimits}=100$ and ${\mathop{\rm Re}\nolimits}=200$.}
            \label{fig:Fd-Fl-S-100-200}
        \end{figure}

\begin{figure}[H]
            \centering
            \subfigure[Streamlines and velocity contours for $Re=100$]{
                \label{fig:stream-re=100}
                \includegraphics[width=7cm]{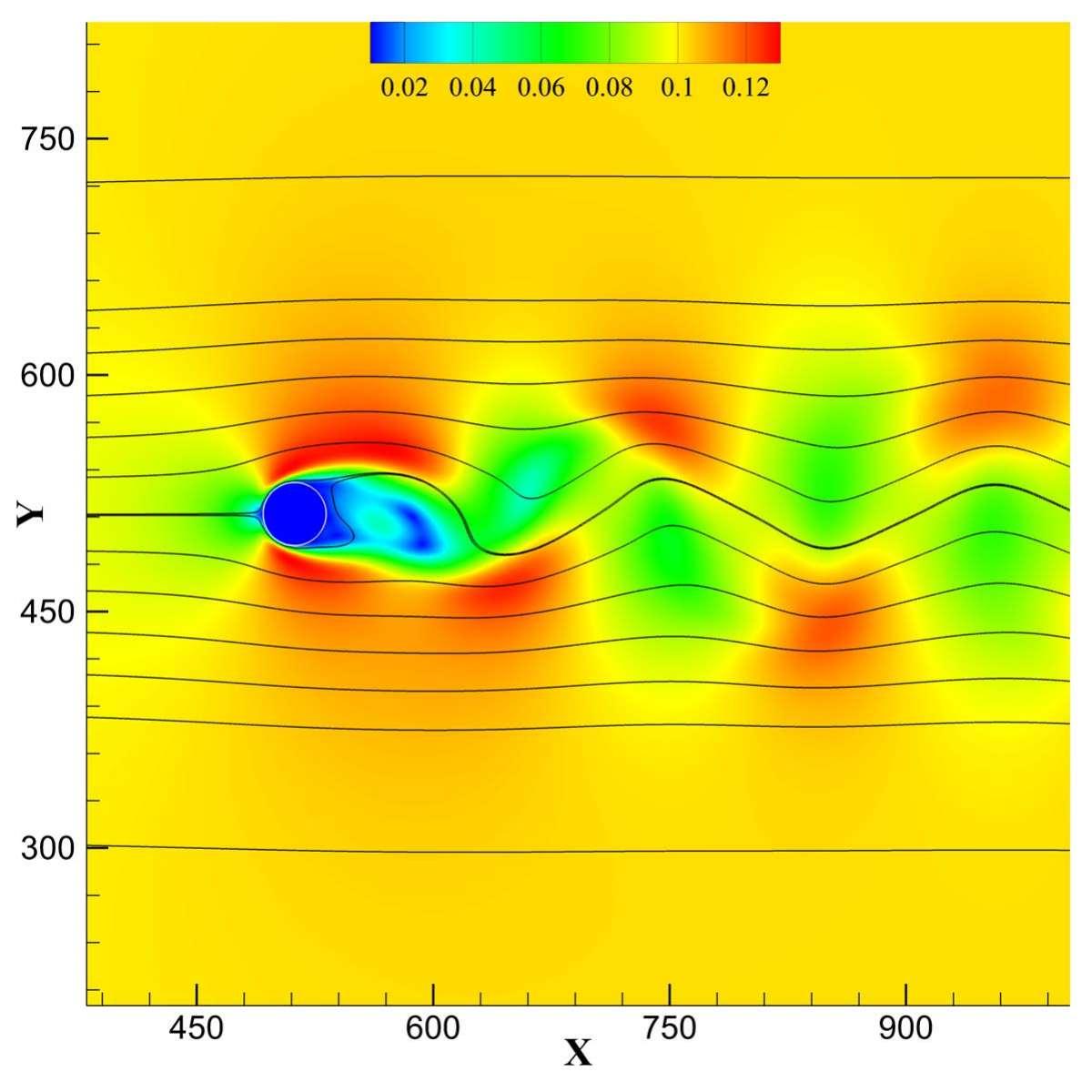}}
            \subfigure[Streamlines and velocity contours for $Re=200$]{
                \label{fig:stream-re=200}
                \includegraphics[width=7cm]{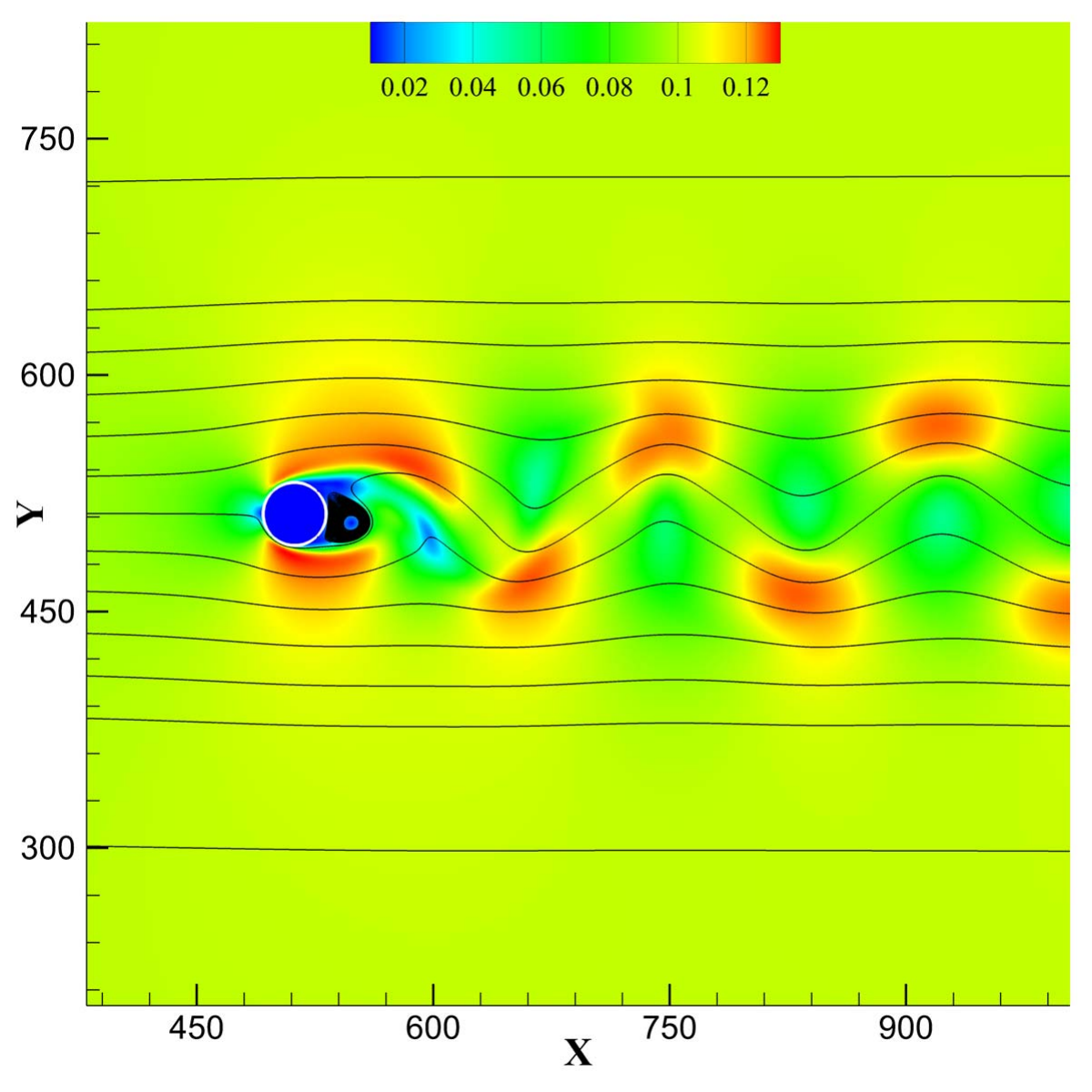}}
                \captionsetup{justification=centering}
            \caption{Unsteady flow past a cylinder at ${\mathop{\rm Re}\nolimits}=100$ and ${\mathop{\rm Re}\nolimits}=200$.}
            \label{fig:figure.high-Re}
        \end{figure}

        When the Reynolds number turns to $100$ and $200$, the flows become unsteady. A famous phenomenon will occur: a K\'{a}rm\'{a}n vortex street develops, as shown in Fig. \ref{fig:figure.high-Re}. As a result of vortex shedding, a lift will also act on the cylinder. And the drag and lift coefficient will also not remain unchanged, which is shown in Fig. \ref{fig:Fd-Fl-S-100-200}. Not only the drag and lift coefficient but also the Strouhal number will be studied and compared with the results in the previous literatures. From Table \ref{tab:high-Re}, the present results agree well with the data in the previous literatures. The evolutions of drag and lift coefficients are presented in Fig. \ref{fig:Fd-Fl-S-100-200}, and the streamlines and velocity contours in Fig. \ref{fig:figure.high-Re}.
        \begin{table}[H]
            \centering
                \captionsetup{justification=centering}
            \caption{Drag, lift coefficients and Strouhal number for flows past a cylinder at ${\mathop{\rm Re}\nolimits}  = 100$ and ${\mathop{\rm Re}\nolimits}  = 200$.}
            \begin{tabular}{p{1.7cm}<{\centering}|p{4.4cm}<{\centering}|p{2.5cm}<{\centering}|p{2.5cm}<{\centering}|p{2.5cm}<{\centering}}
              \hline
              Case & References & ${C_d}$ & ${C_l}$ & ${S_t}$ \\
              \hline \multirow{4}{*}{${\mathop{\rm Re}\nolimits}  = 100$} & Benson et al. \cite{Benson-1989} & $1.46 \pm 0.01$ & $ \pm 0.38$ & $0.17$\\
              \cline{2-5} & Ding et al. \cite{Ding-Shu-2004} & $1.325 \pm 0.008$ & $ \pm 0.28$ & $0.164$ \\
              \cline{2-5} & J. Wu and C. Shu \cite{Wu-Shu-I-V} & $1.334 \pm 0.012$ & $\pm 0.37$ & $0.163$ \\
              \cline{2-5} & Present & $1.336 \pm 0.008$ & $\pm 0.304$ & $0.169$ \\
              \hline \multirow{4}{*}{${\mathop{\rm Re}\nolimits}  = 200$} & Benson et al. \cite{Benson-1989} & $1.45 \pm 0.04$ & $\pm 0.65$ & $0.193$ \\
              \cline{2-5} & Ding et al. \cite{Ding-Shu-2004} & $1.327 \pm 0.045$ & $\pm 0.60$ & $0.196$ \\
              \cline{2-5} & J. Wu and C. Shu \cite{Wu-Shu-I-V} & $1.43 \pm 0.051$ & $\pm 0.75$ & $0.195$\\
              \cline{2-5} & Present & $1.355 \pm 0.040$ & $ \pm 0.638$ & $ 0.196 $ \\
              \hline
            \end{tabular}
            \label{tab:high-Re}
        \end{table}

    \subsection{\label{sec:two-cylinder} Flows past a pair of circular cylinders}
\begin{figure}[H]
            \centering
            \subfigure[Tandem]{
                \label{fig:tandem-arrange}
                \includegraphics[width=7cm]{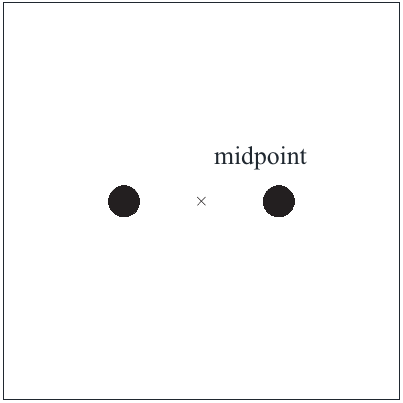}}
            \subfigure[Side-by-side]{
                \label{fig:side-by-side-arrange}
                \includegraphics[width=7cm]{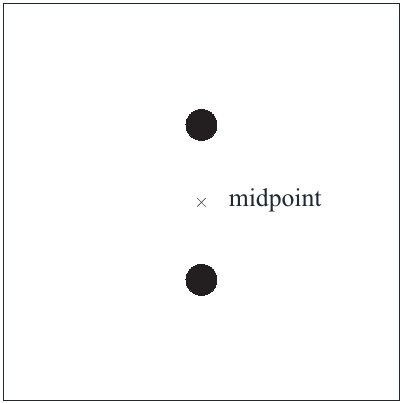}}
                \captionsetup{justification=centering}
            \caption{Tandem and Side-by-side arrangements.}
            \label{fig:two-type-arrange}
        \end{figure}

        Compared with the flows past a single cylinder, the flows past a pair of circular cylinders are some more complicated, which are conducted to validate the presented method's capacity of solving the problems involving complex flows. As same as in the previous literatures, side-by-side and tandem arrangements, in which the two circular cylinders are set, are chosen as the experiments, as shown in the Fig. \ref{fig:two-type-arrange}. In all experiments, the midpoint between the two cylinders' centers is located at $(H/2,H/2)$. In the experiments flows past a pair of cylinders, not only the Reynolds number but also the distance between the two cylinders $Lg$ influences the flow fields. In this article, four experiments with different distances $(Lg/D=1.5,2.0,3.0,4.0)$ are conduced for each arrangement. As same as in the previous literatures \cite{Zdravkovich-1977, Igarashi-1981, MENEGHINI2001327, HU2014140}, the Reynolds number is set as $Re=200$. The drag and lift coefficients are taken into comparison with the results in previous literatures, as well as the Strouhal numbers in some cases.
        \subsubsection{\label{sec:tandem-cylinder} Tandem arrangements}

\begin{figure}[H]
                    \centering
                    \subfigure[$Lg=1.5D$]{
                        \label{fig:STT.a}
                        \includegraphics[width=7cm]{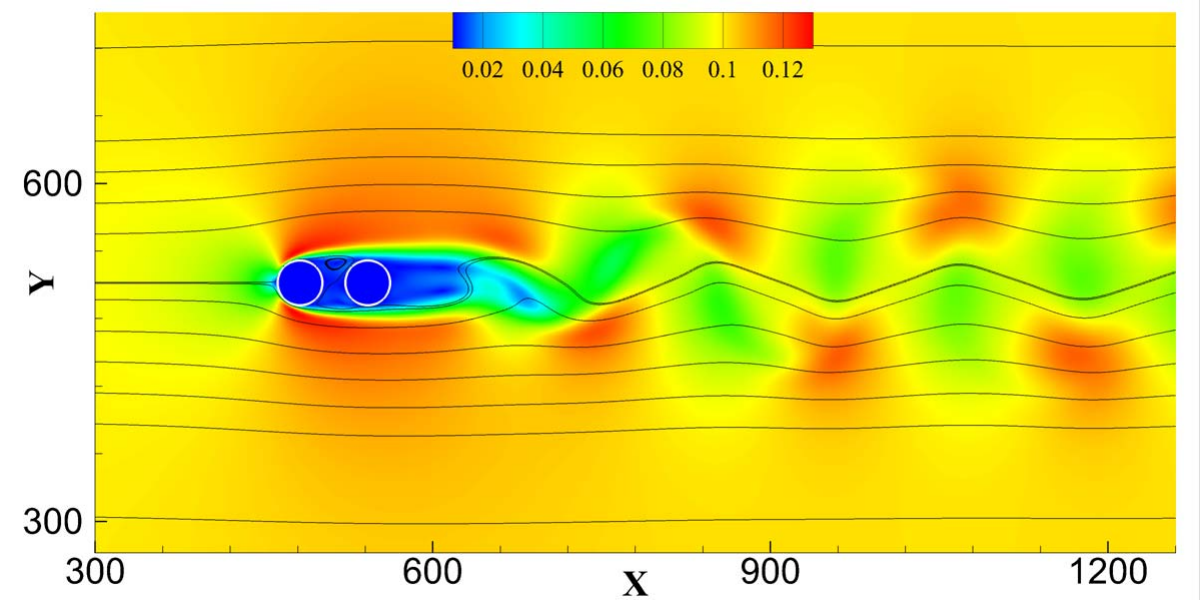}}
                    \subfigure[$Lg=2.0D$]{
                        \label{fig:STT.b}
                        \includegraphics[width=7cm]{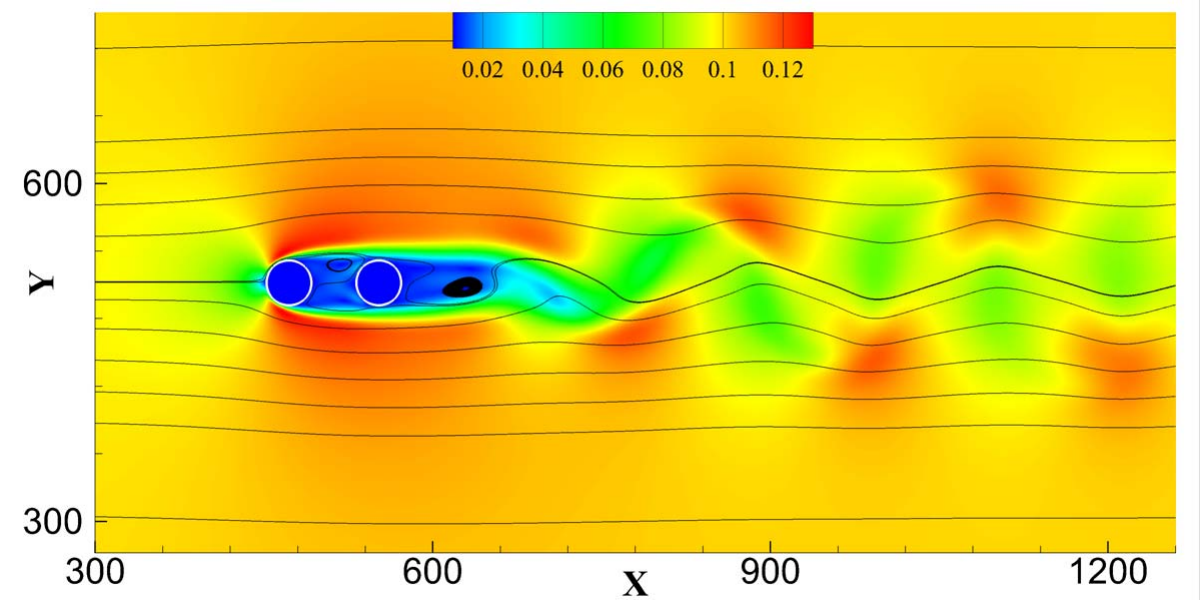}}
                    \subfigure[$Lg=3.0D$]{
                        \label{fig:STT.c}
                        \includegraphics[width=7cm]{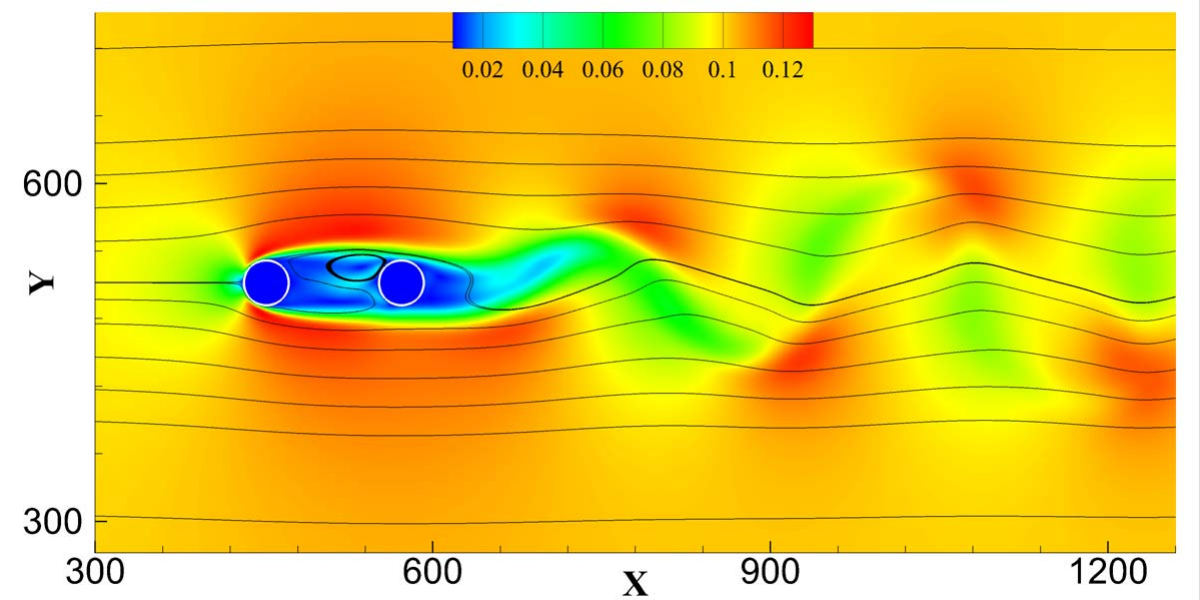}}
                    \subfigure[$Lg=4.0D$]{
                        \label{fig:STT.d}
                        \includegraphics[width=7cm]{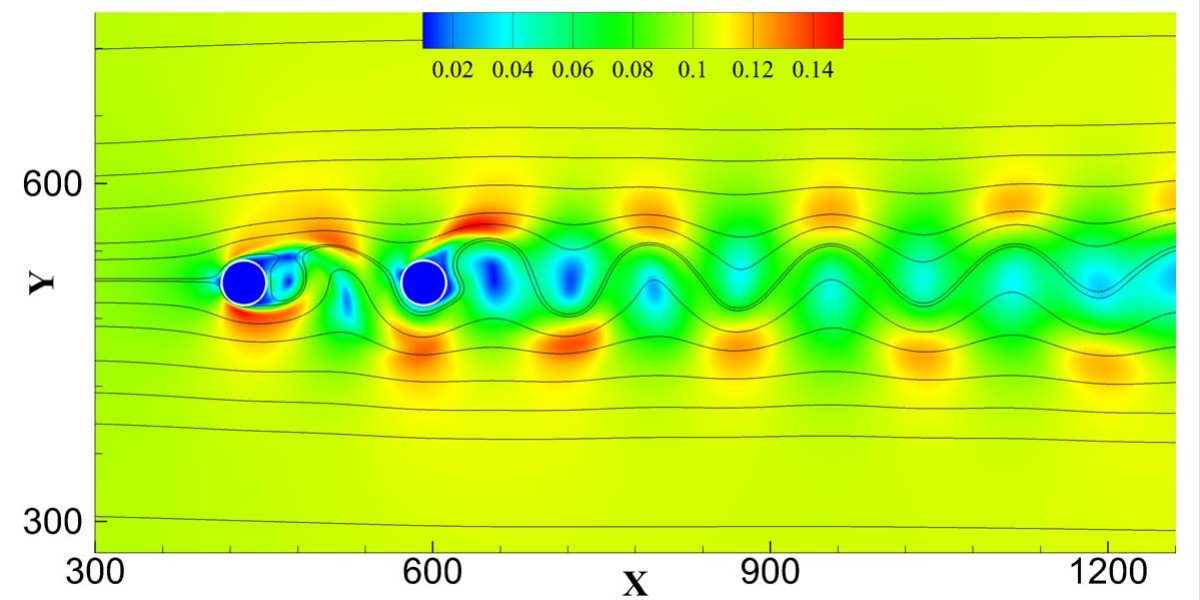}}
                        \captionsetup{justification=centering}
                    \caption{ Streamline and velocity contours for flows past two cylinders in tandem arrangement.}
                    \label{fig:stream-two-tandem}
            \end{figure}

\begin{figure}[H]
                \centering
                \subfigure[$Lg=1.5D$]{
                    \label{fig:FDTT.a}
                    \includegraphics[width=7cm]{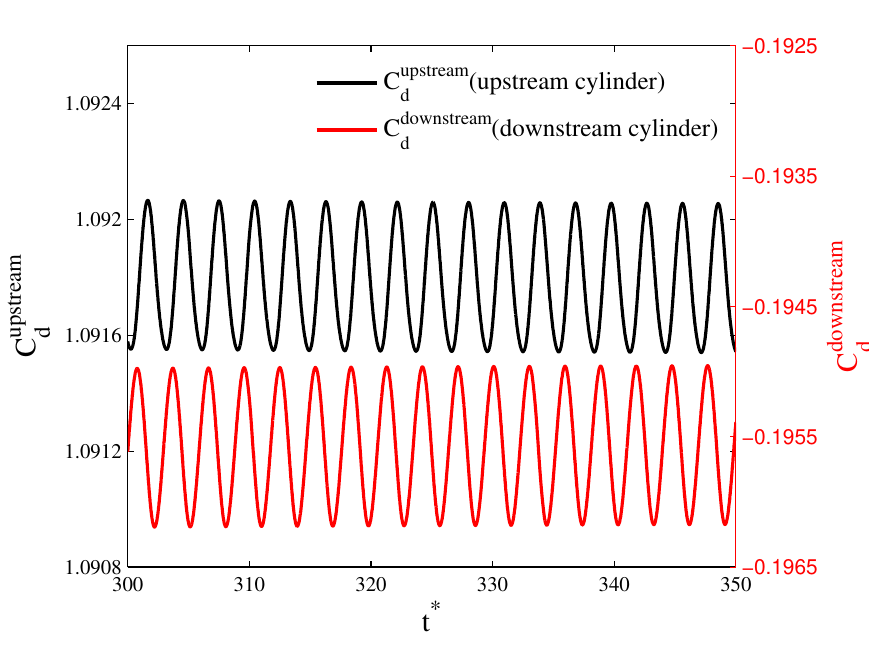}}
                \subfigure[$Lg=2.0D$]{
                    \label{fig:FDTT.b}
                    \includegraphics[width=7cm]{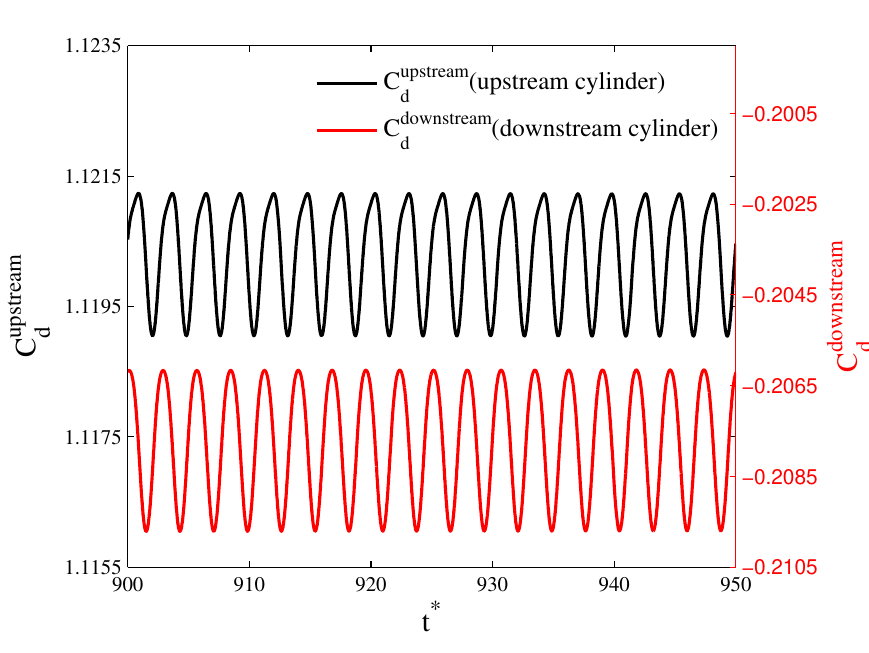}}
                \subfigure[$Lg=3.0D$]{
                    \label{fig:FDTT.c}
                    \includegraphics[width=7cm]{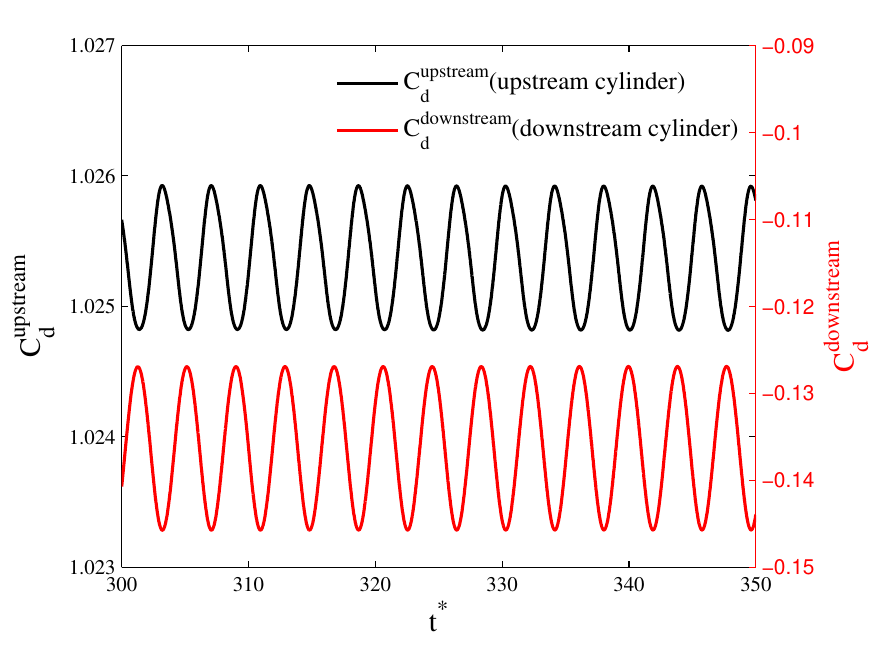}}
                \subfigure[$Lg=4.0D$]{
                    \label{fig:FDTT.d}
                    \includegraphics[width=7cm]{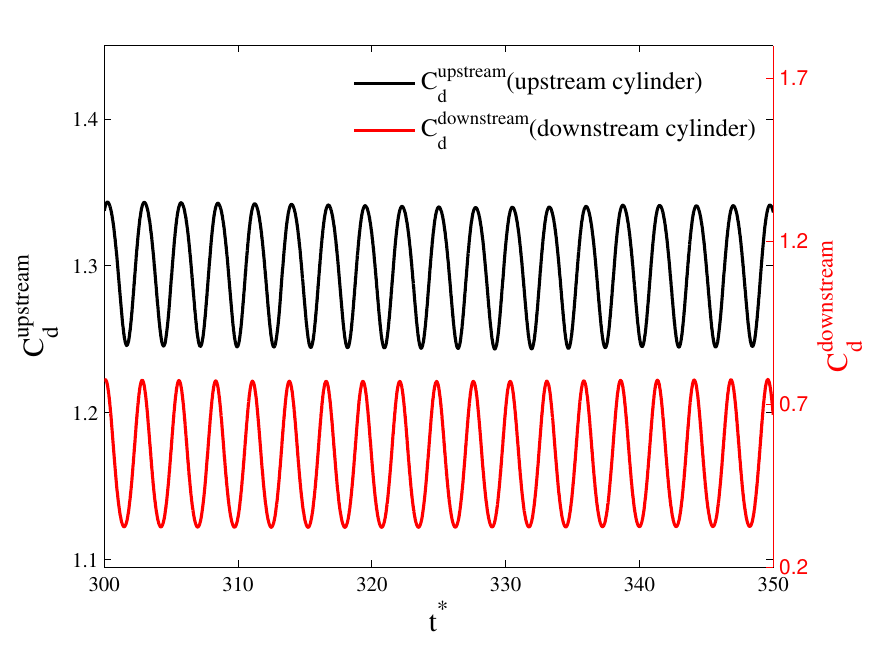}}
                    \captionsetup{justification=centering}
                \caption{ Evolution of drag coefficients of two cylinders in tandem arrangement. }
                \label{fig:FD-two-tandem}
            \end{figure}

            From Fig. \ref{fig:stream-two-tandem}, vortex shedding develops behind the downstream cylinder, as same as in the flows past a single cylinder. However, only when the distance $Lg$ is equal to $4$, vortex shedding occurs between the two cylinders. And the drag coefficients of the upstream cylinders are positive, while the downstream cylinders' are negative, as shown in Fig.  \ref{fig:FD-two-tandem}. For each experiment, drag coefficients of both cylinders oscillate at a same frequency, meaning the Strouhal numbers are equal. The comparison of the results are shown in Tab. \ref{tab:FD-two-tandem}, from which a good agreement can be obtained.

            \begin{table}[H]
                \newcommand{\tabincell}[2]{\begin{tabular}{@{}#1@{}}#2\end{tabular}}
                \centering
                    \captionsetup{justification=centering}
                \caption{The time-average Drag coefficient and Strouhal for flows past a pair of circular cylinders in tandem arrangement at $Re=200$}
                \begin{tabular}{p{1.7cm}<{\centering}|p{3.5cm}<{\centering}|p{1.5cm}<{\centering}|p{1.5cm}<{\centering}|p{1.5cm}<{\centering}|p{1.5cm}<{\centering}}
                    \hline
                        \multicolumn{2}{c|}{$\textup{Lg/D}$}& $1.5$ & $2.0$ & $3.0$ & $4.0$ \\
                      \hline \multirow{3}{*}{\tabincell{c}{Upstream \\ Cylinder}} & Meneghini et al. \cite{MENEGHINI2001327} & $1.06$ & $1.03$ & $1.00$ & $1.18$\\
                      \cline{2-6} & Hu et al. \cite{HU2014140} & $1.158$ & $1.126$ & $1.080$ & $1.355$ \\
                      \cline{2-6} & Present & $1.091$ & $1.120$ & $1.025$ & $1.297$\\
                      \hline \multirow{3}{*}{\tabincell{c}{Downstream \\ Cylinder}} & Meneghini et al. \cite{MENEGHINI2001327}  & $-0.18$ & $-0.17$ & $-0.08$ & $0.38$ \\
                      \cline{2-6} &Hu et al. \cite{HU2014140} & $-0.197$ & $-0.209$ & $-0.140$ & $0.582$ \\
                      \cline{2-6} & Present & $-0.196$ & $-0.208$ & $-0.136$ & $0.529$ \\
                      \hline \multirow{3}{*}{\tabincell{c}{Strouhal \\ Number}} & Meneghini et al. \cite{MENEGHINI2001327} & $0.167$ & $0.130$ & $0.125$ & $0.174$\\
                      \cline{2-6} & Hu et al. \cite{HU2014140} & $0.175$ & $0.171$ & $0.127$ & $0.179$ \\
                      \cline{2-6} & Present & $0.171$ & $0.179$ & $0.129$ & $0.179$\\
                    \hline
                \end{tabular}
                \label{tab:FD-two-tandem}
            \end{table}
        \subsubsection{\label{sec:sbs-cylinder} Side-by-side arrangements}

\begin{figure}[H]
                    \centering
                    \subfigure[$Lg=1.5D$]{
                        \label{fig:STsbs.a}
                        \includegraphics[width=7cm]{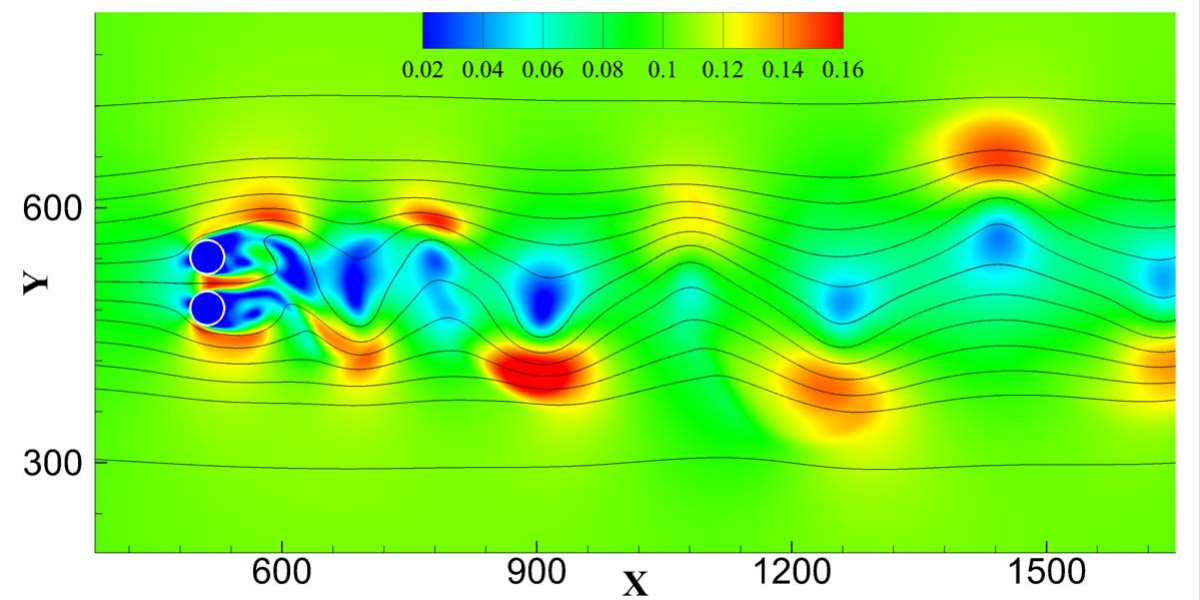}}
                    \subfigure[$Lg=2.0D$]{
                        \label{fig:STsbs.b}
                        \includegraphics[width=7cm]{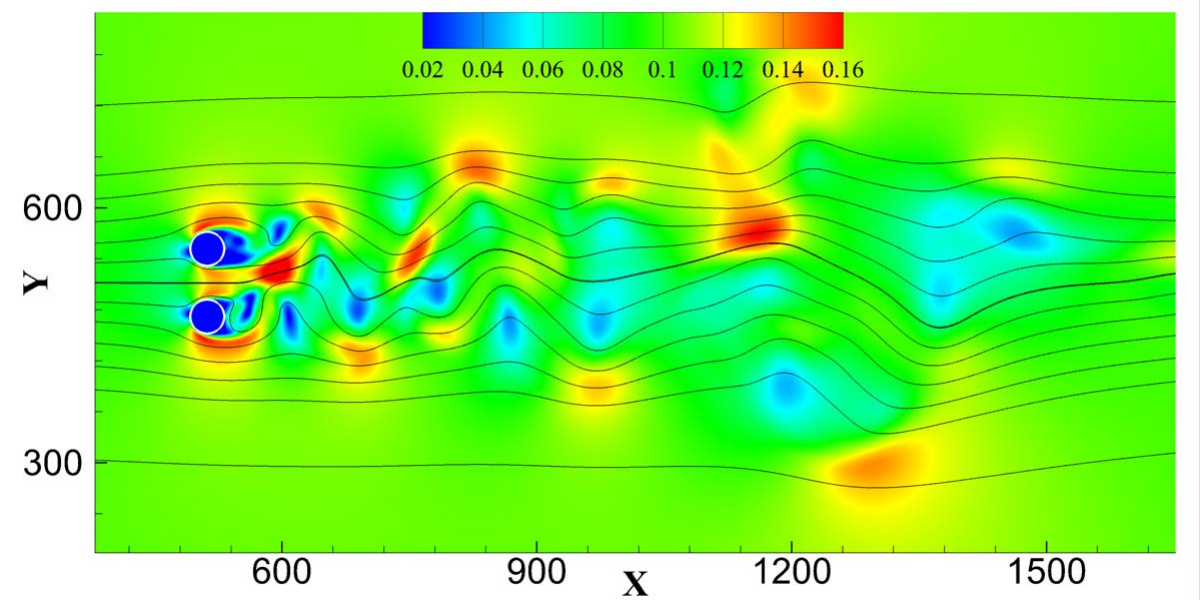}}
                    \subfigure[$Lg=3.0D$]{
                        \label{fig:STsbs.c}
                        \includegraphics[width=7cm]{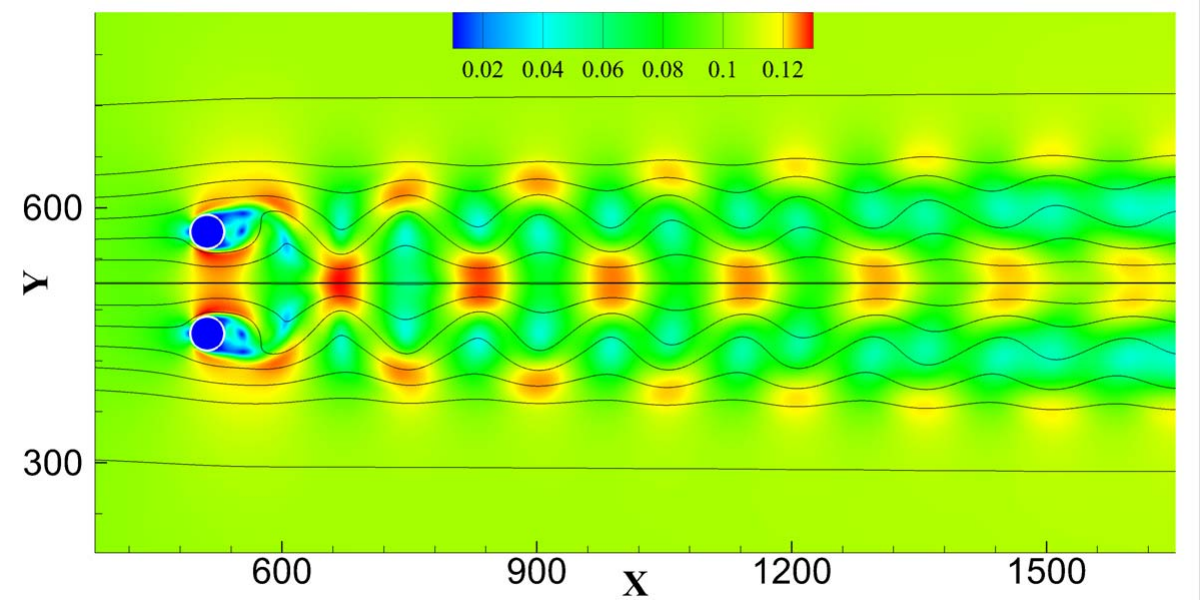}}
                    \subfigure[$Lg=4.0D$]{
                        \label{fig:STsbs.d}
                        \includegraphics[width=7cm]{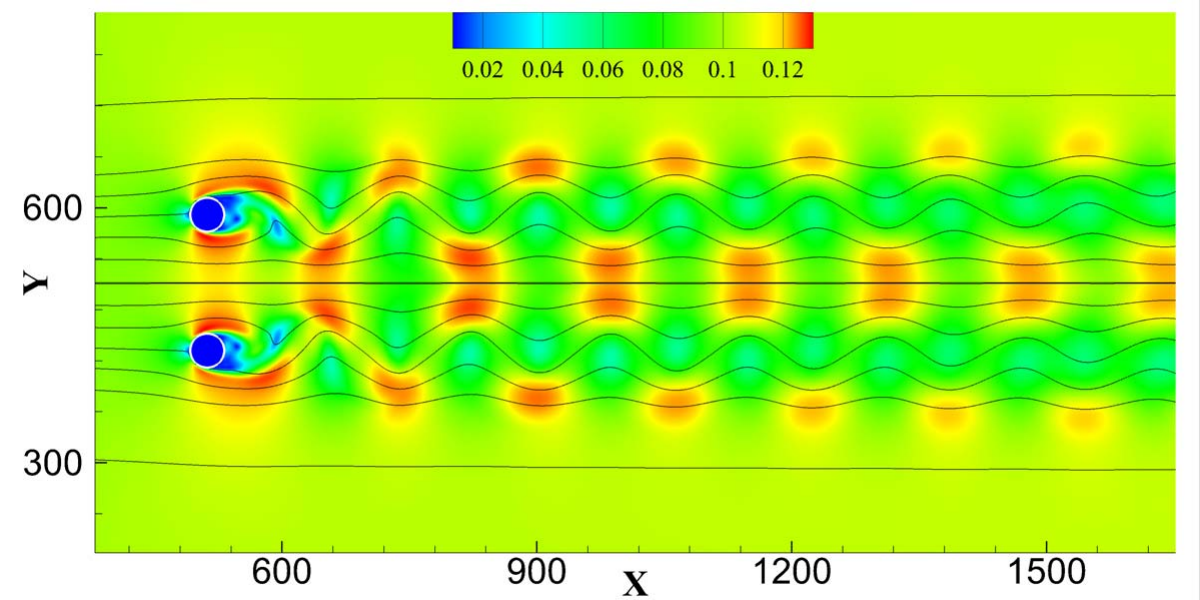}}
                        \captionsetup{justification=centering}
                    \caption{ Streamline and velocity contours for flows past two cylinders in side-by-side arrangement.}
                    \label{fig:stream-two-sbs}
            \end{figure}

\begin{figure}[H]
                \centering
                \subfigure[$Lg=1.5D$]{
                    \label{fig:FDTsbs.a}
                    \includegraphics[width=7cm]{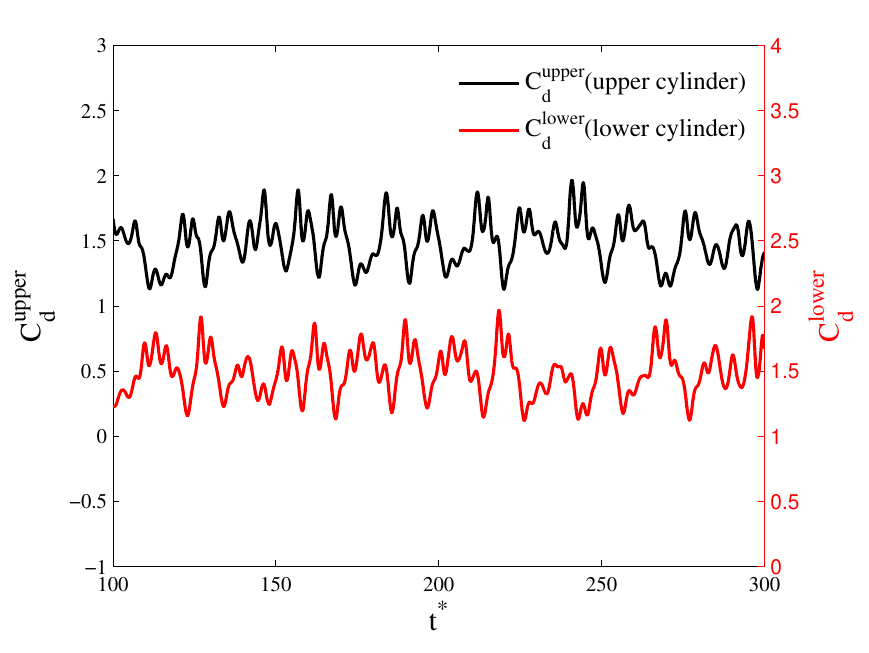}}
                \subfigure[$Lg=2.0D$]{
                    \label{fig:FDTsbs.b}
                    \includegraphics[width=7cm]{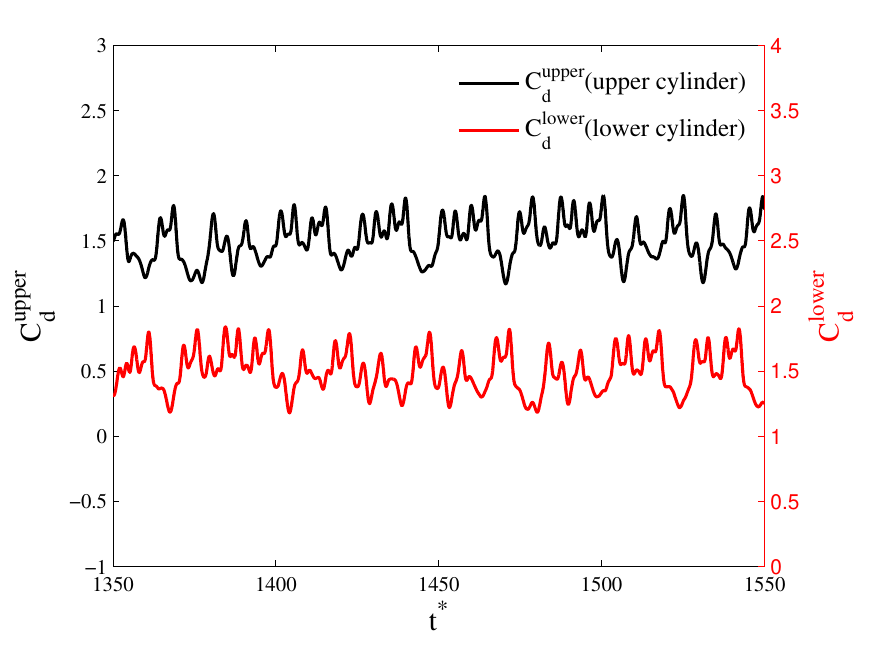}}
                \subfigure[$Lg=3.0D$]{
                    \label{fig:FDTsbs.c}
                    \includegraphics[width=7cm]{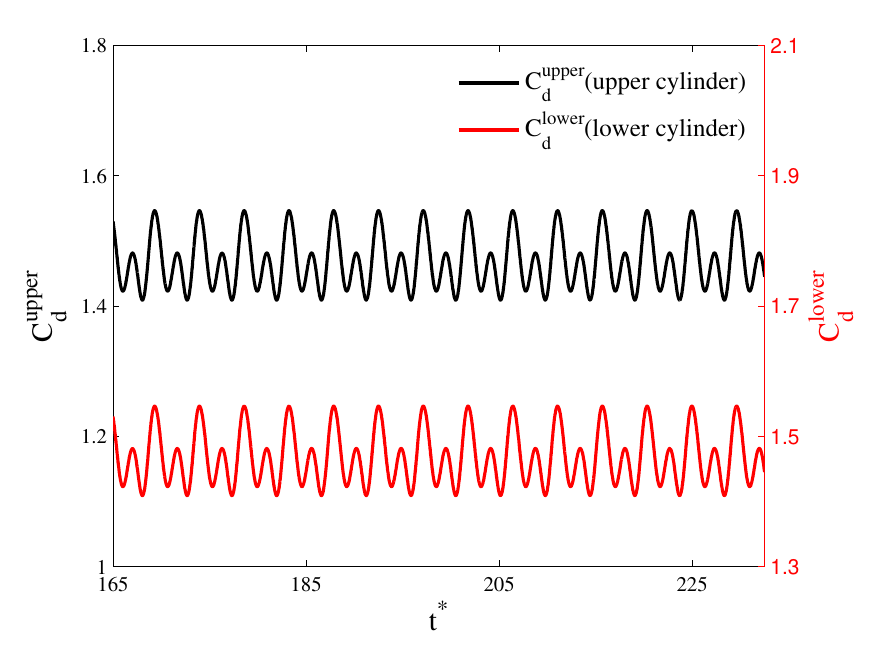}}
                \subfigure[$Lg=4.0D$]{
                    \label{fig:FDTsbs.d}
                    \includegraphics[width=7cm]{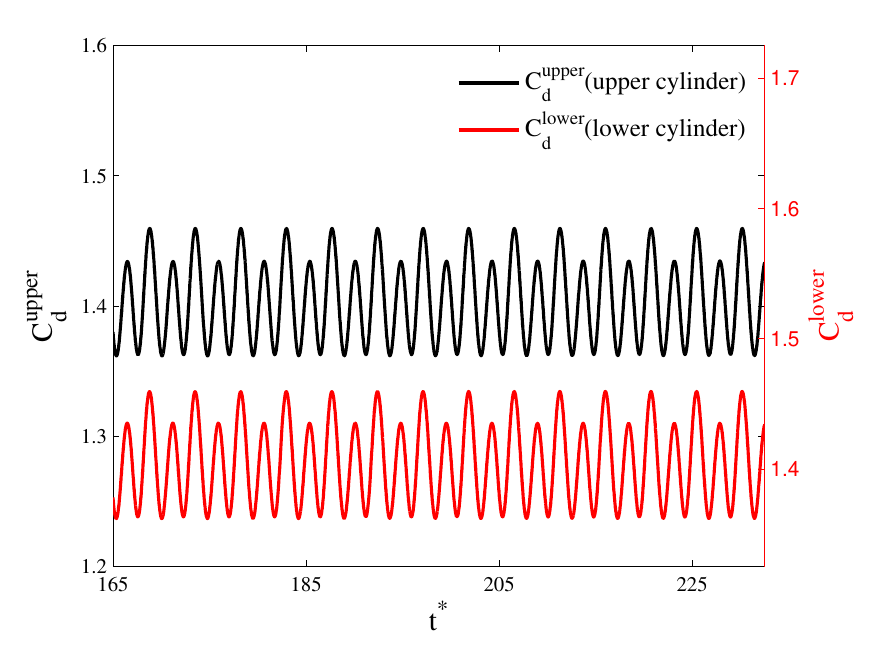}}
                    \captionsetup{justification=centering}
                \caption{ Evolution of drag coefficients of two cylinders in side-by-side arrangement. }
                \label{fig:FD-two-sbs}
            \end{figure}

            It can be seen from Fig. \ref{fig:stream-two-sbs} that there is vortex shedding from the two cylinders for all four different experiments. The flow fields with $Lg=1.5D,2.0D$ are irregular and the wakes of the two experiments are very biased. However, the flow fields of the rest two experiments are of obvious periodicity. And the wakes of both upper and lower cylinders are anti-phase. The velocity contours are symmetric along the horizontal line through the midpoint. From the evolution of the drag coefficients shown in Fig. \ref{fig:FD-two-sbs}, some similar conclusions can be obtained. When the distance $Lg=1.5D,2.0D$, the drag coefficient evolutions shows no obvious cycle but great randomness. However, when the distance $Lg$ turns to $3.0$ and $4.0$, great changes happen. The evolutions of the drag coefficient show periodicity and they oscillate at the same frequency, which means the Strouhal numbers are equal. The time-average drag, lift coefficients and Strouhal numbers are compared with the data in the previous literatures in Tab. \ref{tab:FD-two-sbs}, from which a good agreement can be obtained.

            \begin{table}[H]
                \newcommand{\tabincell}[2]{\begin{tabular}{@{}#1@{}}#2\end{tabular}}
                \centering
                    \captionsetup{justification=centering}
                \caption{The time-average Drag coefficient, lift coefficient and Strouhal number for flows past a pair of circular cylinders in side-by-side arrangement at $Re=200$}
                \begin{tabular}{p{1.7cm}<{\centering}|p{1.5cm}<{\centering}|p{3.5cm}<{\centering}|p{1.5cm}<{\centering}|p{1.5cm}<{\centering}|p{1.5cm}<{\centering}|p{1.5cm}<{\centering}}
                    \hline
                        \multicolumn{3}{c|}{$\textup{Lg/D}$}& $1.5$ & $2.0$ & $3.0$ & $4.0$ \\
                      \hline \multirow{6}{*}{\tabincell{c}{Upper \\ Cylinder}} & \multirowcell{3}{$C_d$} & Meneghini et al. \cite{MENEGHINI2001327} & $1.32$ & $1.42$ & $1.41$ & $1.31$\\
                      \cline{3-7} && Hu et al. \cite{HU2014140} & $1.692$ & $1.718$ & $1.649$ & $1.568$ \\
                      \cline{3-7} && Present & $1.493$ & $1.506$ & $1.468$ & $1.450$\\
                      \cline{2-7}& \multirowcell{3}{$C_l$} & Meneghini et al. \cite{MENEGHINI2001327} & $-0.40$ & $-0.22$ & $-0.10$ & $-0.05$\\
                      \cline{3-7} && Hu et al. \cite{HU2014140} & $-0.490$ & $-0.263$ & $-0.104$ & $-0.052$ \\
                      \cline{3-7} && Present & $-0.441$ & $-0.223$ & $-0.097$ & $-0.054$\\

                      \hline \multirow{6}{*}{\tabincell{c}{Lower \\ Cylinder}} & \multirowcell{3}{$C_d$} & Meneghini et al. \cite{MENEGHINI2001327} & $1.32$ & $1.42$ & $1.41$ & $1.31$\\
                      \cline{3-7} && Hu et al. \cite{HU2014140} & $1.705$ & $1.726$ & $1.649$ & $1.568$ \\
                      \cline{3-7} && Present & $1.490$ & $1.495$ & $1.468$ & $1.425$\\
                      \cline{2-7}& \multirowcell{3}{$C_l$} & Meneghini et al. \cite{MENEGHINI2001327} & $0.40$ & $0.22$ & $0.10$ & $0.05$\\
                      \cline{3-7} && Hu et al. \cite{HU2014140} & $0.518$ & $0.261$ & $0.104$ & $0.052$ \\
                      \cline{3-7} && Present & $0.440$ & $0.228$ & $0.097$ & $0.054$\\

                      \hline \multirow{3}{*}{\tabincell{c}{Both \\ Cylinders}} & \multirowcell{3}{$St$} & Meneghini et al. \cite{MENEGHINI2001327} & $-$ & $-$ & $0.2$ & $0.2$\\
                      \cline{3-7} && Hu et al. \cite{HU2014140} & $-$ & $-$ & $0.214$ & $0.209$ \\
                      \cline{3-7} && Present & $-$ & $-$ & $0.215$ & $0.211$\\

                    \hline

                \end{tabular}
                \label{tab:FD-two-sbs}
            \end{table}
        \subsection{\label{sec:two-cylinder-rotating} Flows past two counter-rotating cylinders}

\begin{figure}[H]
                \centering
                \includegraphics[width=4cm]{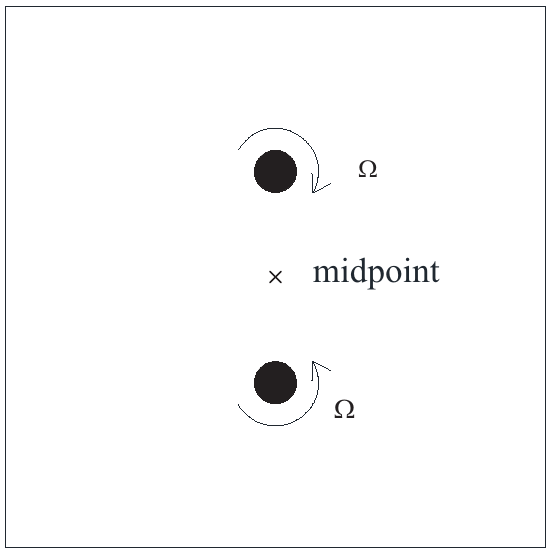}
                \caption{Schematic diagram for flows past two counter-ratating cylinders}\label{fig:Cd-rotating}
            \end{figure}

            The flows past two counter-rotating cylinders is of great interest. Due to the counter-rotation, the flow patterns are very different to those of the flows past two fixed cylinders \cite{Wang2015,chan2010suppression,Chan2011Vortex}. Just as same as in the experiments of flows past a pair of cylinders arranged in side-by-side arrangement, the midpoint between the two cylinders' centers is located at $(H/2,H/2)$ as shown in Fig. \ref{fig:Cd-rotating}. The two cylinders rotate with a same normalized angular velocity $\omega  = {{\Omega D} \mathord{\left/
             {\vphantom {{\Omega D} {\left( {2{U_\infty }} \right)}}} \right. \kern-\nulldelimiterspace} {\left( {2{U_\infty }} \right)}}$. The distance between the two cylinders $Lg$ are equal to $2D$, the Reynolds number $Re=150$ and the free stream velocity ${U_\infty}=0.01$. Six experiments with $\omega=1.5,2.0,2.5,3.0,3.5,4.0$ are carried out in this article.

\begin{figure}[htbp]
                    \centering
                    \subfigure[$\omega=1.5$]{
                        \label{fig:w=1.5}
                        \includegraphics[width=7cm]{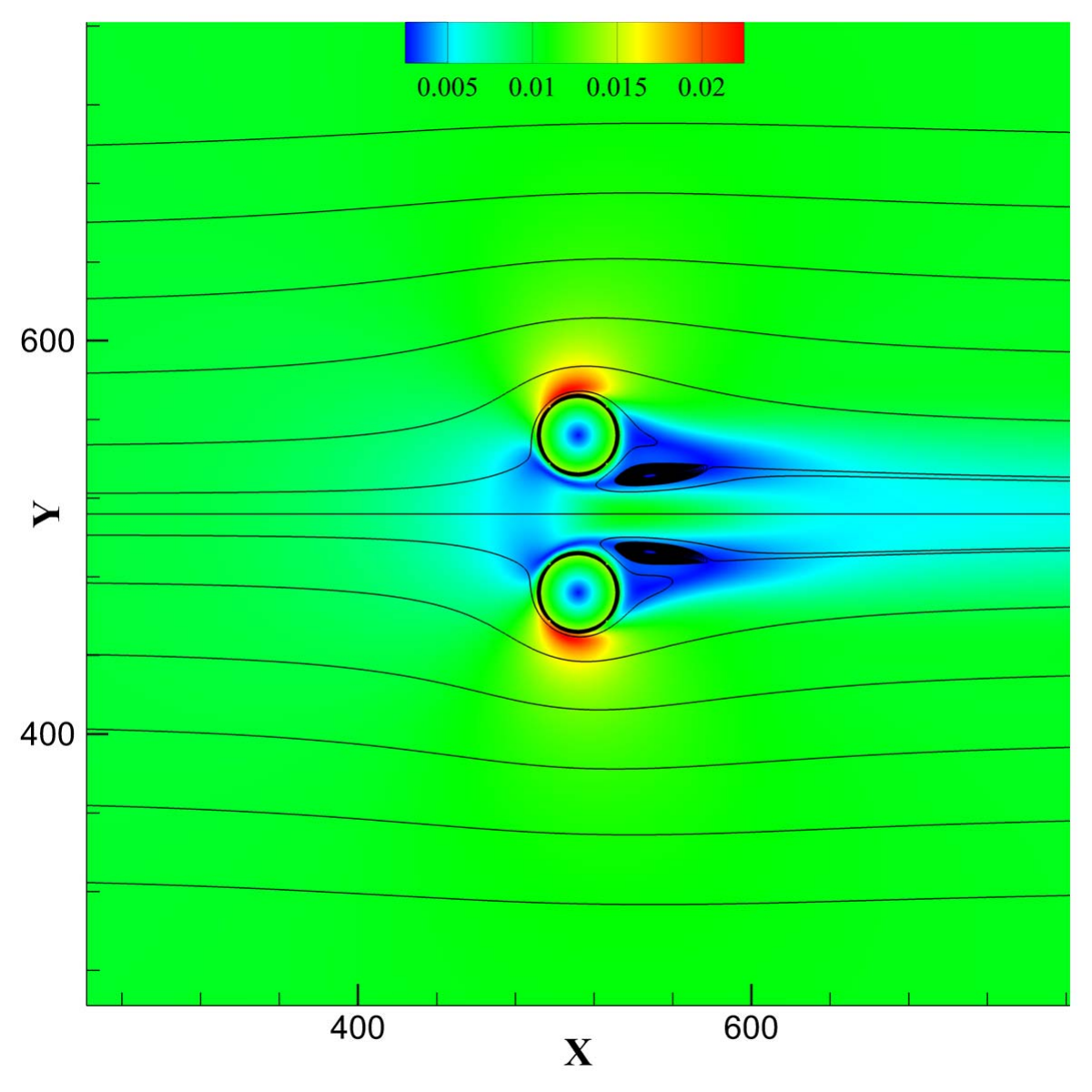}}
                    \subfigure[$\omega=2.0$]{
                        \label{fig:w=2.0}
                        \includegraphics[width=7cm]{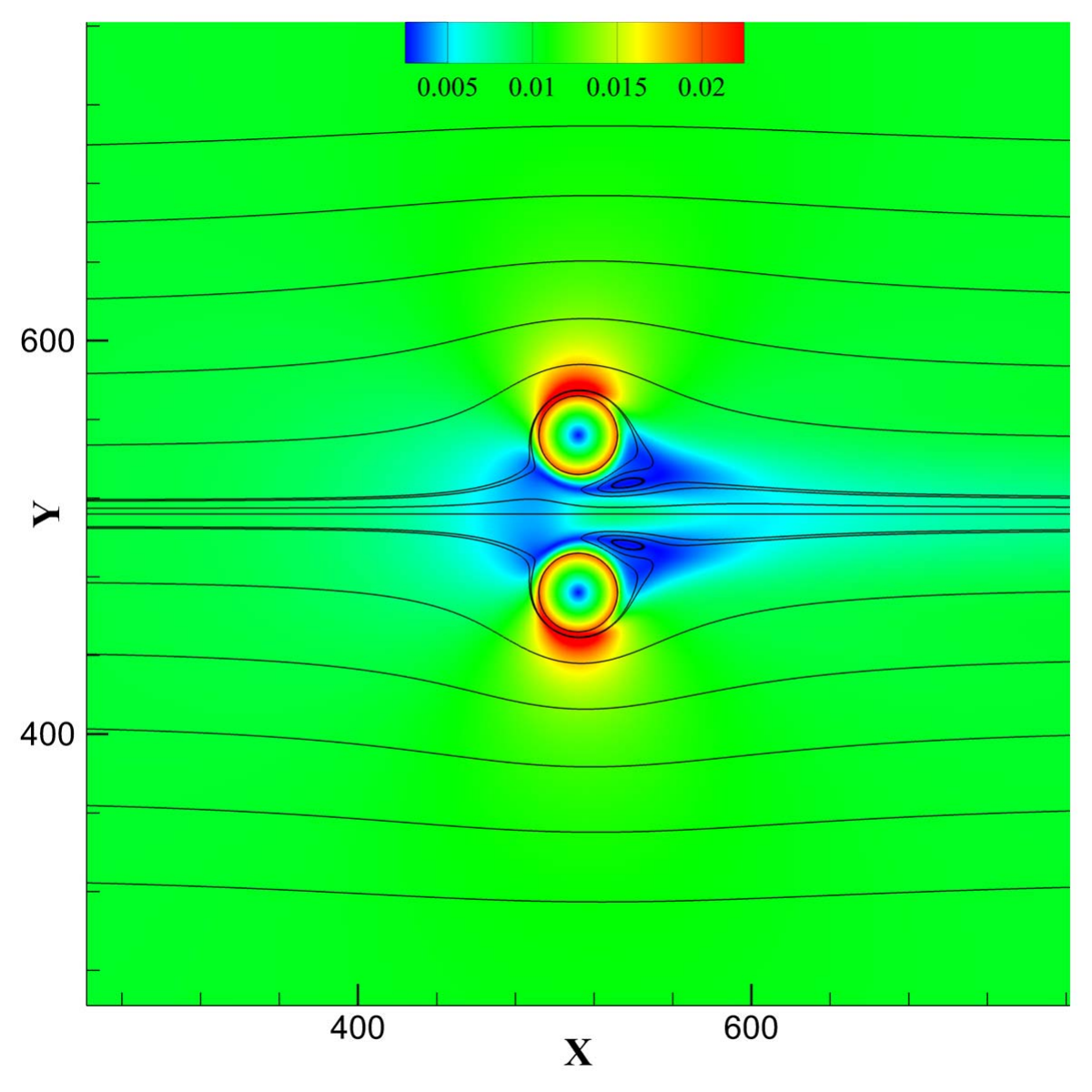}}
                    \subfigure[$\omega=2.5$]{
                        \label{fig:w=2.5}
                        \includegraphics[width=7cm]{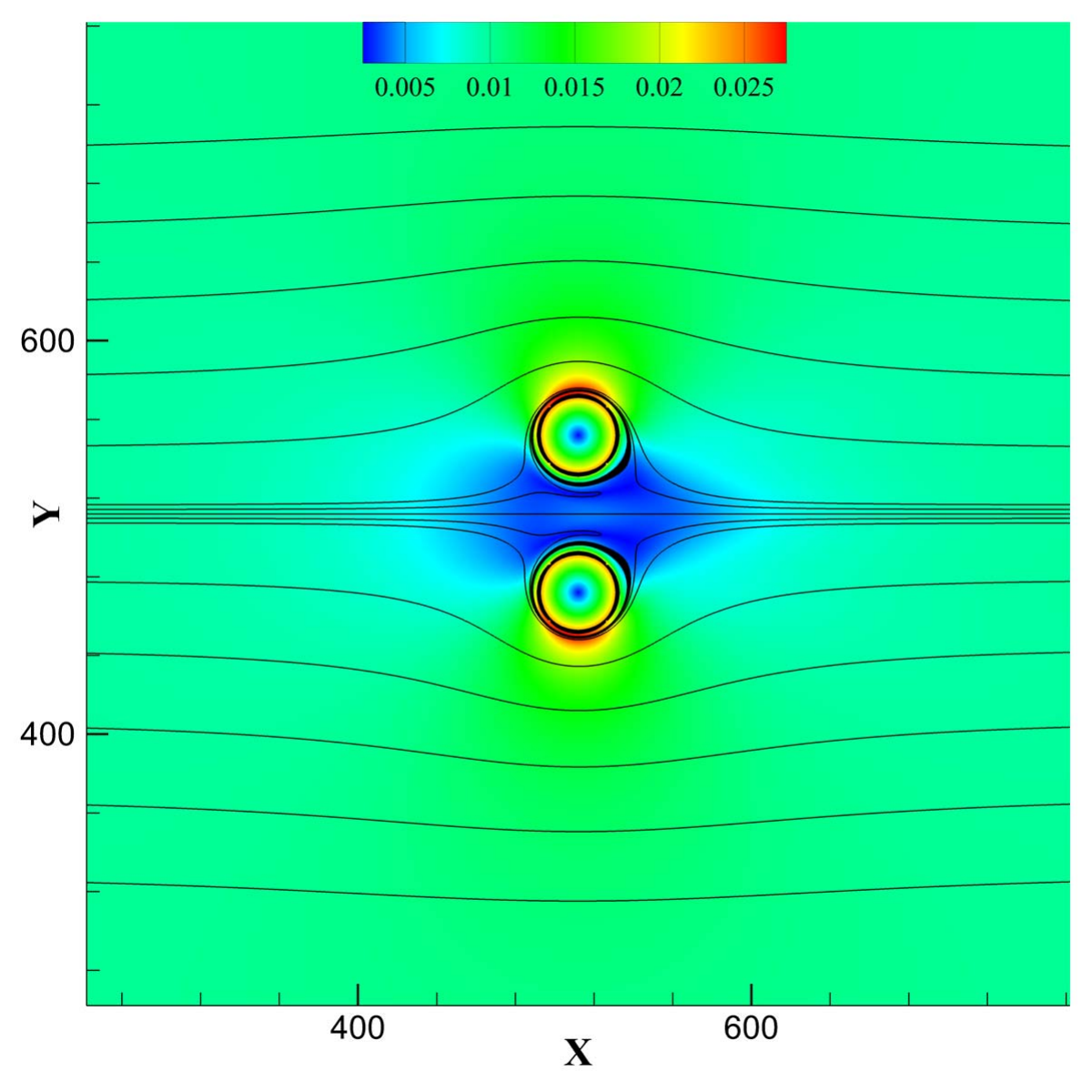}}
                    \subfigure[$\omega=3.0$]{
                        \label{fig:w=3.0}
                        \includegraphics[width=7cm]{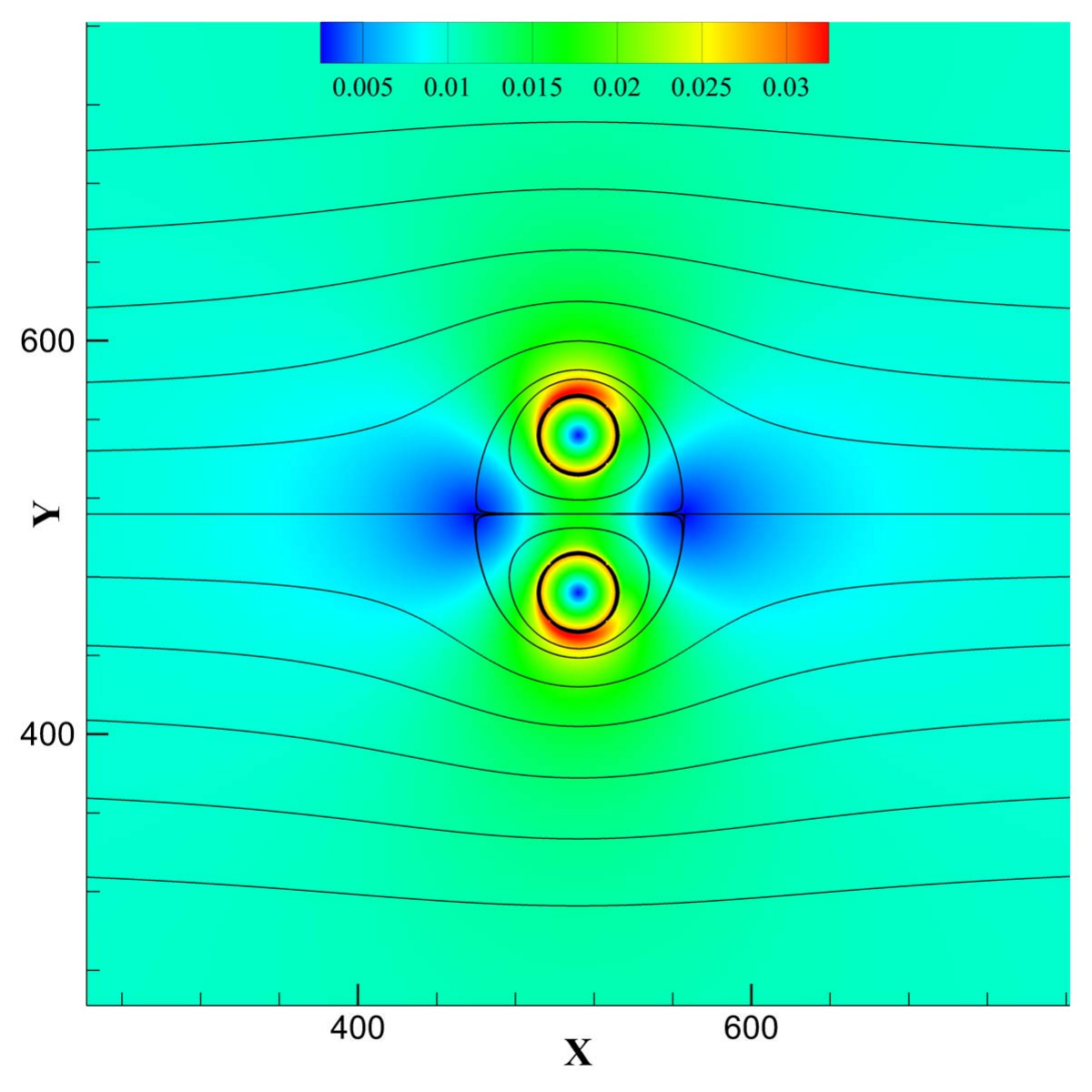}}
                    \subfigure[$\omega=3.5$]{
                        \label{fig:w=3.5}
                        \includegraphics[width=7cm]{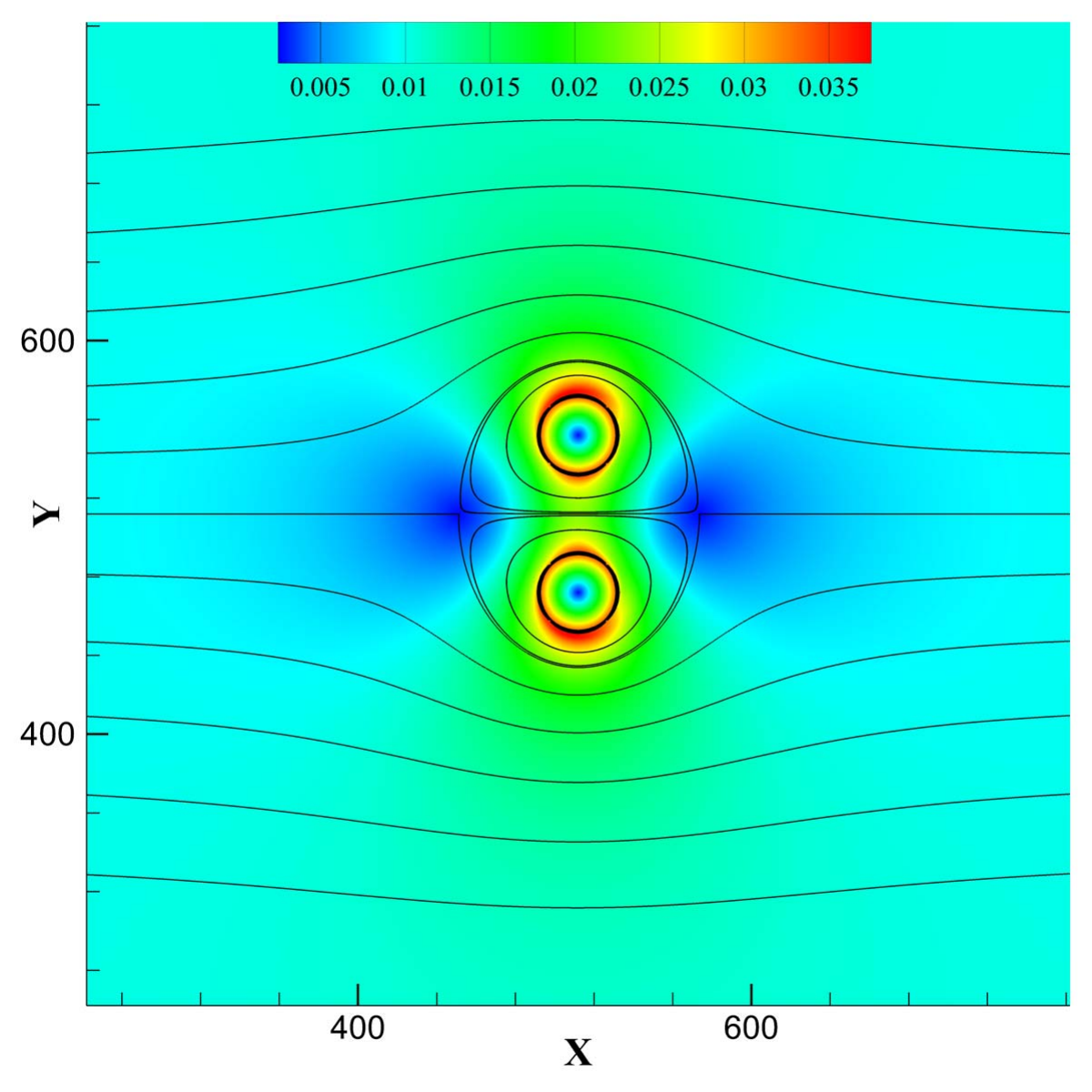}}
                    \subfigure[$\omega=4.0$]{
                        \label{fig:w=4.0}
                        \includegraphics[width=7cm]{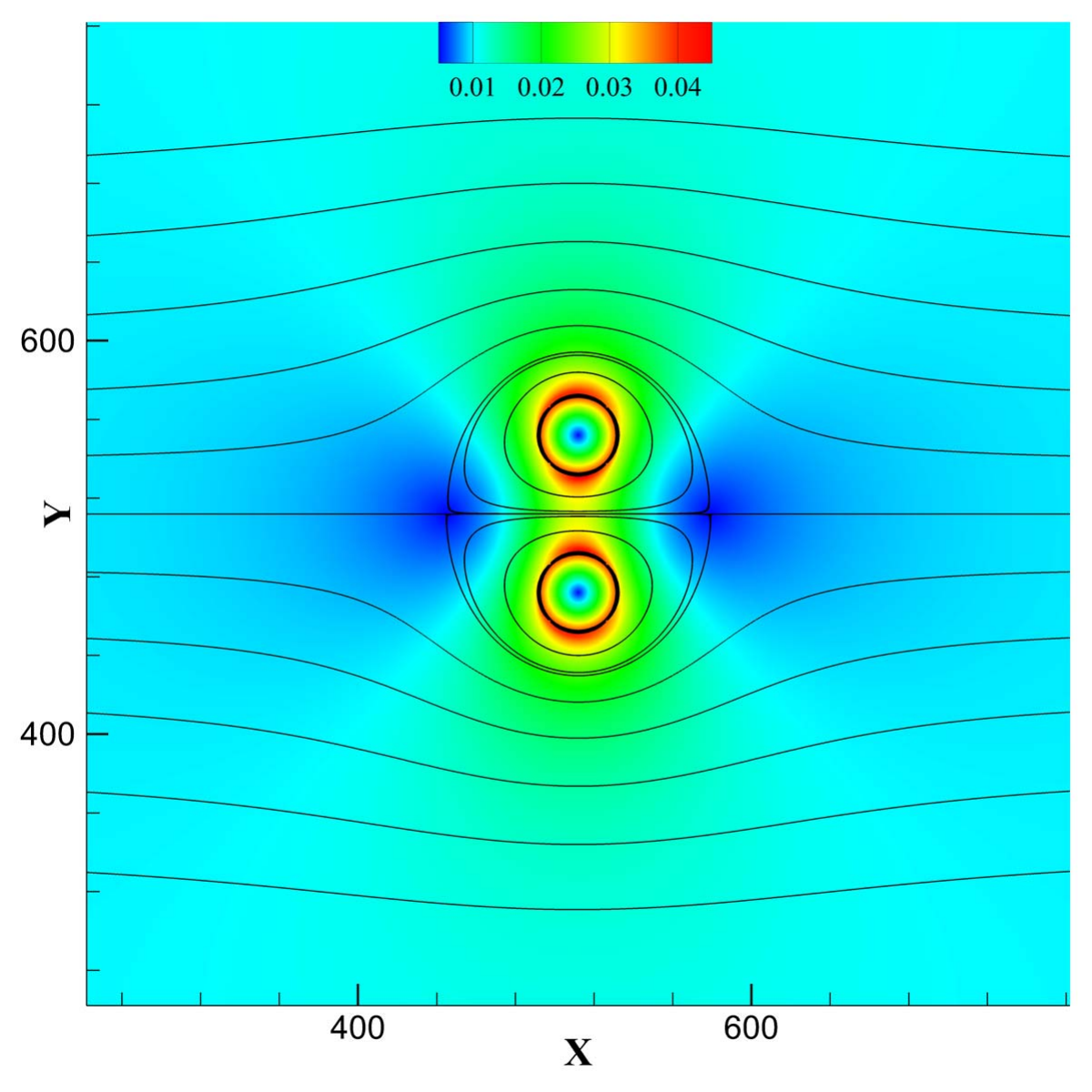}}
                    \captionsetup{justification=centering}
                    \caption{ Streamlines and velocity contours for flows past two counter-rotating cylinders.}
                    \label{fig:stream-two-rotating}
            \end{figure}

\begin{figure}[htbp]
                    \centering
                    \subfigure{
                        \includegraphics[width=7cm]{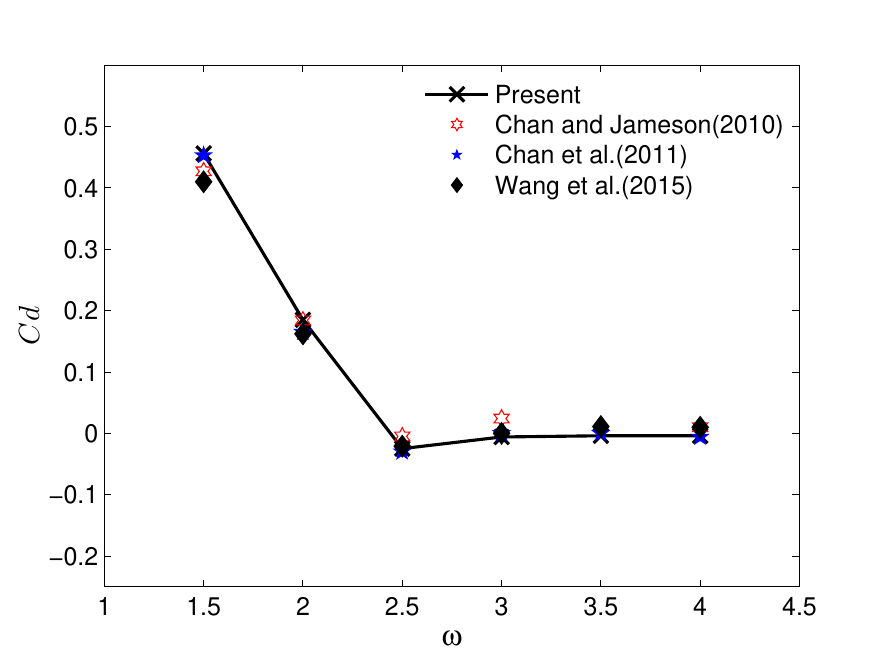}}
                    \subfigure{
                        \includegraphics[width=7cm]{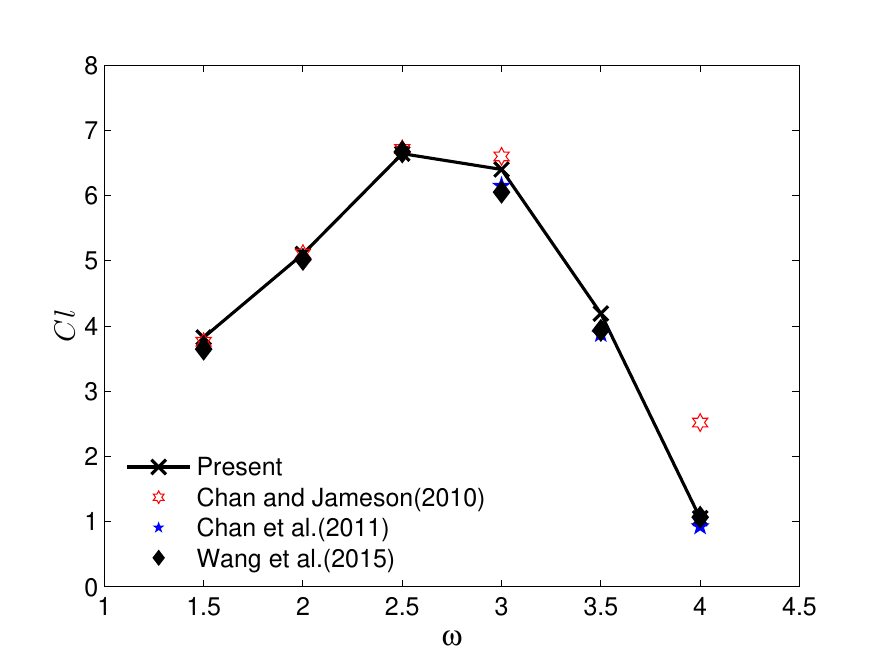}}
                    \captionsetup{justification=centering}
                    \caption{Comparisons of the drag and lifte coefficients of the lower cylinder for flows past two counter-rotating cylinders with $Re=150$.}
                    \label{fig:CdCl-two-rotating}
            \end{figure}

            The flow patterns change according to the normalized rotational speed $\omega$. As pointed in the previous literatures \cite{chan2010suppression,Chan2011Vortex,Wang2015}, the unsteady wakes behind the two cylinders can be completely suppressed with $\omega>1.5$ and the flow fields are symmetric along the horizontal line through the midpoint, which can be observed in the present results shown in Fig \ref{fig:stream-two-rotating}. Besides, there is no vortex shedding. For the cases of $\omega=1.5$ and $2.0$, the vortex is limited to a certain region near the cylinders. No vortex shedding occurs, which is very different from the flows past two fixed cylinders in the same arrangement. When $\omega=3.0,3.5$ and $4.0$, a virtual elliptic body is observed around the two cylinders. From the streamlines in Fig. \ref{fig:w=3.0}, Fig. \ref{fig:w=3.5} and Fig. \ref{fig:w=4.0}, it can be found that the flow outside the elliptic body do not penetrate the elliptic boundary and the flow inside do not goes outside as well. Just as there was an impenetrable "wall". In the Fig. \ref{fig:CdCl-two-rotating}, the drag and lift coefficients of the lower cylinders are shown. The drag coefficient are reduced significantly when the rotational angle velocity $\omega$ grows from $1.5$ to $2.5$. However, the lift coefficient is firstly increased. When $\omega$ turns from $2.5$ to $4.0$ it is decreased. In the Fig. \ref{fig:CdCl-two-rotating}, the results of Chan and Jameson \cite{chan2010suppression}, Chan et al. \cite{Chan2011Vortex} and Wang \cite{Wang2015} are also figured. Obviously, our results agree well with the others' results.

        \subsection{\label{sec:airfoil} Flows past a NACA-0012 airfoil}
            For the practical application of the TRT-LBM-VP method, the flows past a NACA-0012 airfoil at $Re=500$ with $AOA=0^\circ$ and at $Re=1000$ with $AOA=10^\circ$ are chosen as the numerical experiments. The free stream velocity is $0.1$ and the chord of the airfoil is $320$.

            \begin{figure}[htbp]
                \centering
                \includegraphics[width=7cm]{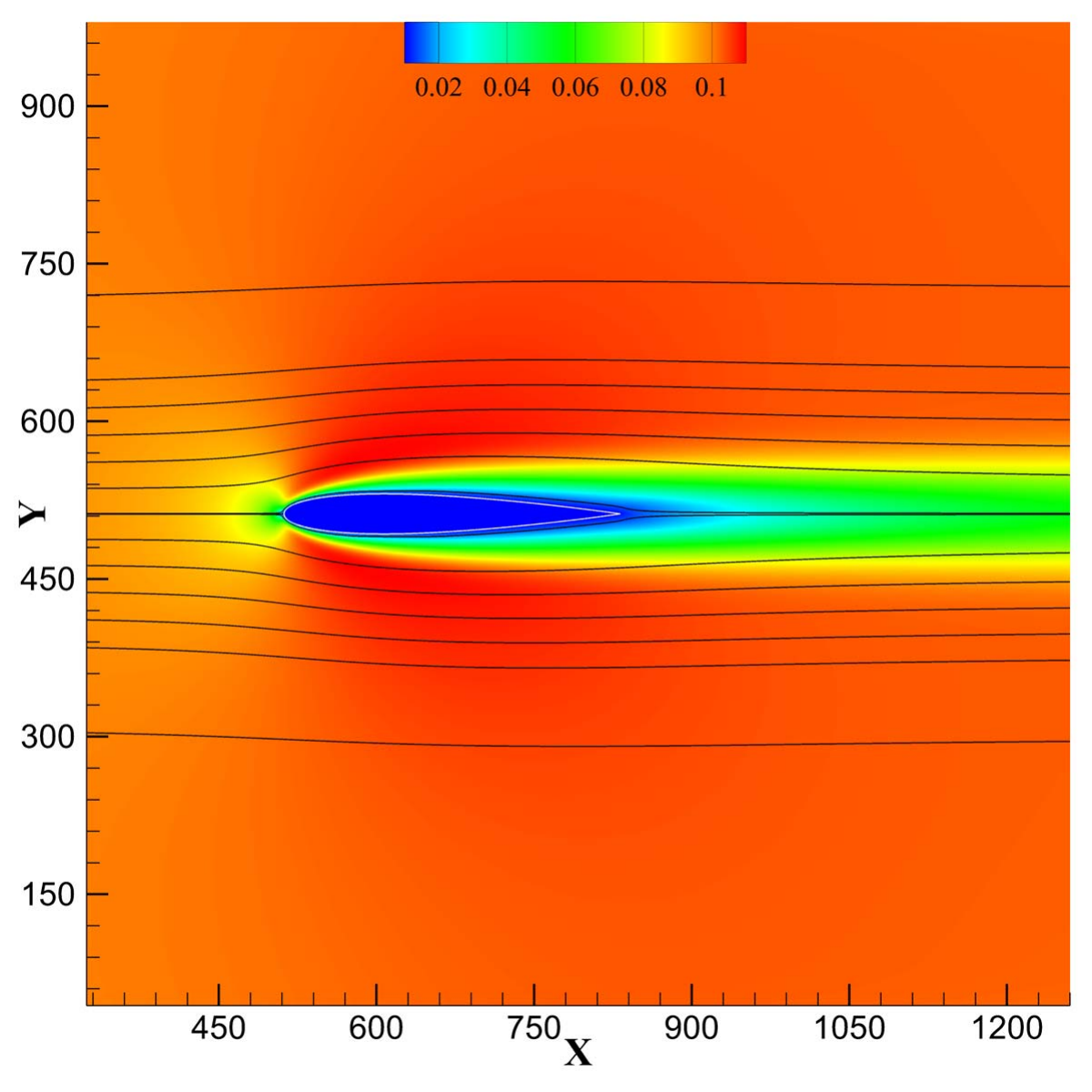}
                \caption{Streamlines and velocity contours for flows past a NACA-0012 airfoil at $Re=500$ with $AOA=0^\circ$}\label{fig:NACA-500-0}
            \end{figure}

\begin{figure}[htbp]
                \centering
                \includegraphics[width=7cm]{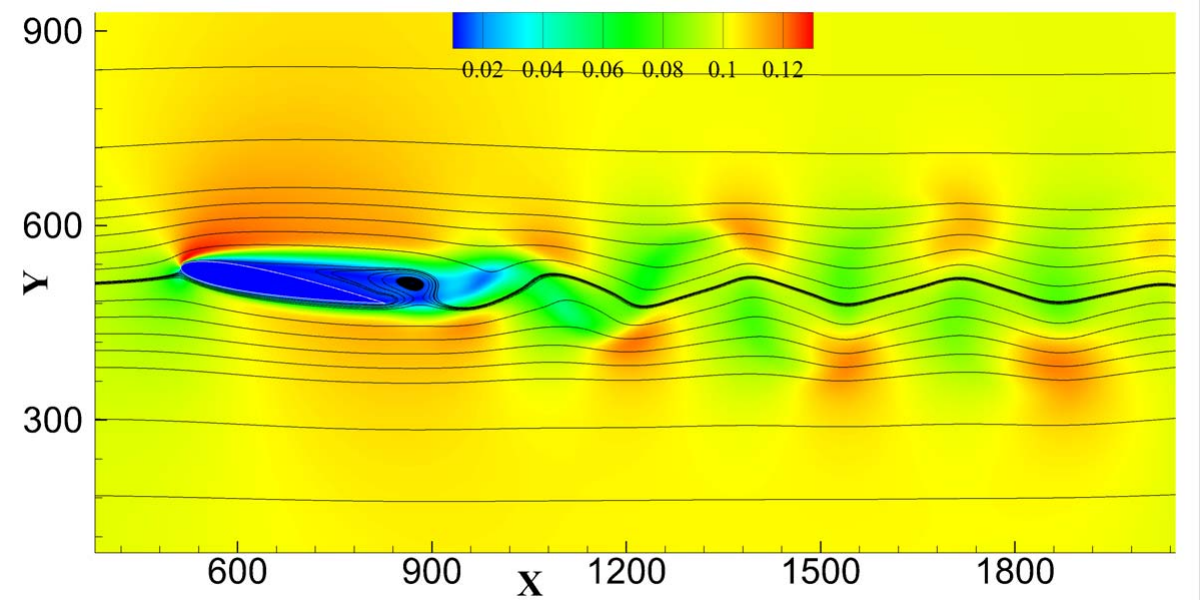}
                \caption{Streamlines and velocity contours for flows past a NACA-0012 airfoil at $Re=1000$ with $AOA=10^\circ$}\label{fig:NACA-1000-10}
            \end{figure}

\begin{figure}[H]
                \centering
                \includegraphics[width=7cm]{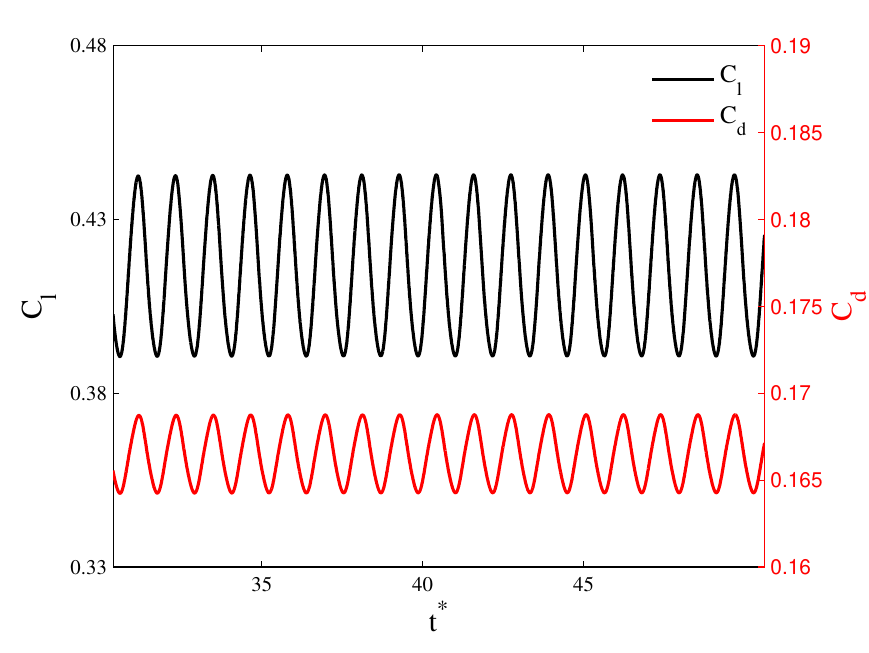}
                \caption{Drag and lift coefficients for flows past a NACA-0012 airfoil at $Re=1000$ with $AOA=10^\circ$}\label{fig:FDCL-NACA-1000-10}
            \end{figure}

            The streamlines and velocity contours of the flows past a NACA-0012 airfoil are shown in Fig. \ref{fig:NACA-500-0} and Fig. \ref{fig:NACA-1000-10}. No vortex shedding occurs behind the airfoil at $Re=500$ with $AOA=0^\circ$, and a steady flow field develops. The drag coefficient of the airfoil is $0.178$, agreeing well with the results obtained by Lockard et al. \cite{Lockard2002} with 0.1762 and Wu et al. \cite{Wu-Shu-I-V} with 0.1759. For the case of $Re=1000$ with $AOA=10^\circ$, vortex shedding occurs. As a result, there is a lift force acting on the airfoil. In the present experiment, the Strouhal number for the drag coefficient is $0.8633$, while the value obtained by Falagkaris et al. \cite{FALAGKARIS20172348} is 0.861 and by Mittal and Tezduyar \cite{MITTAL1994253} is $0.862$. The drag and lift coefficients are figured in Fig. \ref{fig:FDCL-NACA-1000-10}.

\section{\label{sec:Con}Conclusions}
    In this article, the Volume Penalization method is incorporated into the Two-relaxation-time Lattice Boltzmann method to simulate the flows past obstacles. Through the force term proposed by Guo et al. \cite{Guo2002-LBM-force}, the force contributed from the boundary ais introduced into the LBM and then divided into symmetrical and antisymmetrical parts. In the TRT-LBM, the collision operator and the force term are relaxed by two relaxation times, which are related to each other by a magic parameter. Compared to the direct-forcing and velocity correction IBM, there is no need to interpolate the velocity at boundaries by using a delta function and to distribute the force density to the Eulerian point near the boundary. However, on a certain Lagrangian point, only the variables of one certain Euler point are needed in the VP procedure, meaning the TRT-LBM-VP can be conducted parallelly. Firstly, the cylindrical Couette flow experiments using different grid size solved by the TRT-LBM-VP and the SRT-LBM-VP are conducted to compare the accuracy of these two methods. The results show that the accuracy of the TRT-LBM-VP is higher than that of the SRT-LBM-VP. To justify the present method further, flows past a single circular cylinder with $Re=20,40,100$ and $200$ and some slightly more complex experiments: flows past a pair of circular cylinder arranged in tandem and side-by-side arrangements, flows past two counter-rotating cylinders are studied. For the practical application, flows past a NACA-0012 airfoil at $Re=500$ with $AOA=0^\circ$ and at $Re=1000$ with $AOA=10^\circ$ are selected as experiments.

    From the comparisons between the present numerical results and the data in the previous literatures, good agreements are achieved. Although the current method is developed and justified in 2D, it is easy to extend the present method to 3D by replacing D2Q9 lattice model with D3Q15 or D3Q19 lattice model.

\section{Acknowledgements}

The authors would like to acknowledge the support of  the National Nature Science Foundation of China (52001094, 51879048, 52001088) and the support from Weihai Key Laboratory of Fluid Structure Interaction Dynamics.

\section*{References}
\bibliographystyle{elsarticle-num}
\bibliography{MyTRT}

\begin{thebibliography}{10}
\expandafter\ifx\csname url\endcsname\relax
  \def\url#1{\texttt{#1}}\fi
\expandafter\ifx\csname urlprefix\endcsname\relax\def\urlprefix{URL }\fi
\expandafter\ifx\csname href\endcsname\relax
  \def\href#1#2{#2} \def\path#1{#1}\fi

\bibitem{Shiyi1998}
S.~Chen, G.~D. Doolen,
  \href{http://dx.doi.org/10.1146/annurev.fluid.30.1.329}{Lattice boltzmann
  method for fluid flows}, Annual Review of Fluid Mechanics 30~(1) (1998)
  329--364.
\newblock \href
  {http://arxiv.org/abs/http://dx.doi.org/10.1146/annurev.fluid.30.1.329}
  {\path{arXiv:http://dx.doi.org/10.1146/annurev.fluid.30.1.329}}, \href
  {https://doi.org/10.1146/annurev.fluid.30.1.329}
  {\path{doi:10.1146/annurev.fluid.30.1.329}}.
\newline\urlprefix\url{http://dx.doi.org/10.1146/annurev.fluid.30.1.329}

\bibitem{PIQUET2016272}
A.~Piquet, O.~Roussel, A.~Hadjadj,
  \href{http://www.sciencedirect.com/science/article/pii/S0045793016301888}{A
  comparative study of brinkman penalization and direct-forcing immersed
  boundary methods for compressible viscous flows}, Computers \& Fluids
  136~(Supplement C) (2016) 272 -- 284.
\newblock \href
  {https://doi.org/https://doi.org/10.1016/j.compfluid.2016.06.001}
  {\path{doi:https://doi.org/10.1016/j.compfluid.2016.06.001}}.
\newline\urlprefix\url{http://www.sciencedirect.com/science/article/pii/S0045793016301888}

\bibitem{peskin_2002}
C.~S. Peskin, The immersed boundary method, Acta Numerica 11 (2002) 479--517.
\newblock \href {https://doi.org/10.1017/S0962492902000077}
  {\path{doi:10.1017/S0962492902000077}}.

\bibitem{FENG2004602}
E.~E.~M. Zhi-Gang~Feng,
  \href{http://www.sciencedirect.com/science/article/pii/S0021999103005758}{The
  immersed boundary-lattice boltzmann method for solving fluid--particles
  interaction problems}, Journal of Computational Physics 195~(2) (2004) 602 --
  628.
\newblock \href {https://doi.org/https://doi.org/10.1016/j.jcp.2003.10.013}
  {\path{doi:https://doi.org/10.1016/j.jcp.2003.10.013}}.
\newline\urlprefix\url{http://www.sciencedirect.com/science/article/pii/S0021999103005758}

\bibitem{Wu-Shu-I-V}
J.~Wu, C.~Shu,
  \href{http://www.sciencedirect.com/science/article/pii/S0021999108006116}{Implicit
  velocity correction-based immersed boundary-lattice boltzmann method and its
  applications}, Journal of Computational Physics 228~(6) (2009) 1963 -- 1979.
\newblock \href {https://doi.org/http://dx.doi.org/10.1016/j.jcp.2008.11.019}
  {\path{doi:http://dx.doi.org/10.1016/j.jcp.2008.11.019}}.
\newline\urlprefix\url{http://www.sciencedirect.com/science/article/pii/S0021999108006116}

\bibitem{IJM294}
M.~Benamour, E.~Liberge, C.~Beghein,
  \href{http://journal.multiphysics.org/index.php/IJM/article/view/294}{Lattice
  boltzmann method for fluid flow around bodies using volume penalization}, The
  International Journal of Multiphysics 9~(3) (2016).
\newblock \href {https://doi.org/10.1260/1750-9548.9.3.299}
  {\path{doi:10.1260/1750-9548.9.3.299}}.
\newline\urlprefix\url{http://journal.multiphysics.org/index.php/IJM/article/view/294}

\bibitem{Arquis19841}
E.~Arquis, J.~Caltagirone, Sur les conditions hydrodynamiques au voisinage
  d'une interface milieu fluide-milieu poreux: Application a la convection
  naturelle, Comptes Rendus de I'Academie des Sciences Series I - Mathematics
  299~(1) (1984) 1--4, cited By 106.

\bibitem{ango1999}
P.~Angot, C.-H. Bruneau, P.~Fabrie,
  \href{https://doi.org/10.1007/s002110050401}{A penalization method to take
  into account obstacles in incompressible viscous flows}, Numerische
  Mathematik 81~(4) (1999) 49--520.
\newblock \href {https://doi.org/10.1007/s002110050401}
  {\path{doi:10.1007/s002110050401}}.
\newline\urlprefix\url{https://doi.org/10.1007/s002110050401}

\bibitem{KEVLAHAN2001333}
N.~K.-R. Kevlahan, J.-M. Ghidaglia,
  \href{http://www.sciencedirect.com/science/article/pii/S0997754600011213}{Computation
  of turbulent flow past an array of cylinders using a spectral method with
  brinkman penalization}, European Journal of Mechanics - B/Fluids 20~(3)
  (2001) 333 -- 350.
\newblock \href {https://doi.org/https://doi.org/10.1016/S0997-7546(00)01121-3}
  {\path{doi:https://doi.org/10.1016/S0997-7546(00)01121-3}}.
\newline\urlprefix\url{http://www.sciencedirect.com/science/article/pii/S0997754600011213}

\bibitem{SCHNEIDER20051223}
K.~Schneider,
  \href{http://www.sciencedirect.com/science/article/pii/S0045793004001343}{Numerical
  simulation of the transient flow behaviour in chemical reactors using a
  penalisation method}, Computers \& Fluids 34~(10) (2005) 1223 -- 1238.
\newblock \href
  {https://doi.org/https://doi.org/10.1016/j.compfluid.2004.09.006}
  {\path{doi:https://doi.org/10.1016/j.compfluid.2004.09.006}}.
\newline\urlprefix\url{http://www.sciencedirect.com/science/article/pii/S0045793004001343}

\bibitem{PhysRevE.65.046308}
Z.~Guo, C.~Zheng, B.~Shi,
  \href{https://link.aps.org/doi/10.1103/PhysRevE.65.046308}{Discrete lattice
  effects on the forcing term in the lattice boltzmann method}, Physical Review
  E Statistical Nonlinear \& Soft Matter Physics 65 (2002) 046308.
\newblock \href {https://doi.org/10.1103/PhysRevE.65.046308}
  {\path{doi:10.1103/PhysRevE.65.046308}}.
\newline\urlprefix\url{https://link.aps.org/doi/10.1103/PhysRevE.65.046308}

\bibitem{Luo-2011}
L.-S. Luo, W.~Liao, X.~Chen, Y.~Peng, W.~Zhang,
  \href{https://link.aps.org/doi/10.1103/PhysRevE.83.056710}{Numerics of the
  lattice boltzmann method: Effects of collision models on the lattice
  boltzmann simulations}, Physical Review E Statistical Nonlinear \& Soft
  Matter Physics 83 (2011) 056710.
\newblock \href {https://doi.org/10.1103/PhysRevE.83.056710}
  {\path{doi:10.1103/PhysRevE.83.056710}}.
\newline\urlprefix\url{https://link.aps.org/doi/10.1103/PhysRevE.83.056710}

\bibitem{Chen-Doolen-1998}
S.~Chen, G.~D. Doolen,
  \href{https://doi.org/10.1146/annurev.fluid.30.1.329}{Lattice boltzmann
  method for fluid flows}, in: Annual review of fluid mechanics, {V}ol. 30,
  Vol.~30 of Annu. Rev. Fluid Mech., Annual Reviews, Palo Alto, CA, 1998, pp.
  329--364.
\newline\urlprefix\url{https://doi.org/10.1146/annurev.fluid.30.1.329}

\bibitem{Le-Zhang}
G.~Le, J.~Zhang,
  \href{https://link.aps.org/doi/10.1103/PhysRevE.79.026701}{Boundary slip from
  the immersed boundary lattice boltzmann models}, Physical Review E
  Statistical Nonlinear \& Soft Matter Physics 79 (2009) 026701.
\newblock \href {https://doi.org/10.1103/PhysRevE.79.026701}
  {\path{doi:10.1103/PhysRevE.79.026701}}.
\newline\urlprefix\url{https://link.aps.org/doi/10.1103/PhysRevE.79.026701}

\bibitem{MR1902782}
D.~d'Humi\`eres, I.~Ginzburg, M.~Krafczyk, P.~Lallemand, L.-S. Luo,
  \href{https://doi.org/10.1098/rsta.2001.0955}{Multiple-relaxation-time
  lattice {B}oltzmann models in three dimensions}, R. Soc. Lond. Philos. Trans.
  Ser. A Math. Phys. Eng. Sci. 360~(1792) (2002) 437--451, discrete modelling
  and simulation of fluid dynamics (Corse, 2001).
\newline\urlprefix\url{https://doi.org/10.1098/rsta.2001.0955}

\bibitem{Yan2014}
Z.~Yan, M.~Hilpert, \href{https://doi.org/10.1007/s11538-014-0020-1}{A
  multiple-relaxation-time lattice-boltzmann model for bacterial chemotaxis:
  Effects of initial concentration, diffusion, and hydrodynamic dispersion on
  traveling bacterial bands}, Bulletin of Mathematical Biology 76~(10) (2014)
  2449--2475.
\newblock \href {https://doi.org/10.1007/s11538-014-0020-1}
  {\path{doi:10.1007/s11538-014-0020-1}}.
\newline\urlprefix\url{https://doi.org/10.1007/s11538-014-0020-1}

\bibitem{Seta-2014}
T.~Seta, R.~Rojas, K.~Hayashi, A.~Tomiyama,
  \href{https://link.aps.org/doi/10.1103/PhysRevE.89.023307}{Implicit-correction-based
  immersed boundary\char21{}lattice boltzmann method with two relaxation
  times}, Physical Review E Statistical Nonlinear \& Soft Matter Physics 89
  (2014) 023307.
\newblock \href {https://doi.org/10.1103/PhysRevE.89.023307}
  {\path{doi:10.1103/PhysRevE.89.023307}}.
\newline\urlprefix\url{https://link.aps.org/doi/10.1103/PhysRevE.89.023307}

\bibitem{Hayashi-2012}
K.~Hayashi, R.~Rojas, T.~Seta, A.~Tomiyama,
  \href{https://doi.org/10.1260/1757-482X.4.2.193}{Immersed boundary-lattice
  boltzmann method using two relaxation times}, The Journal of Computational
  Multiphase Flows 4~(2) (2012) 193--209.
\newblock \href
  {http://arxiv.org/abs/https://doi.org/10.1260/1757-482X.4.2.193}
  {\path{arXiv:https://doi.org/10.1260/1757-482X.4.2.193}}, \href
  {https://doi.org/10.1260/1757-482X.4.2.193}
  {\path{doi:10.1260/1757-482X.4.2.193}}.
\newline\urlprefix\url{https://doi.org/10.1260/1757-482X.4.2.193}

\bibitem{Luolishi1997}
X.~He, L.-S. Luo,
  \href{http://dx.doi.org/10.1023/B:JOSS.0000015179.12689.e4}{Lattice boltzmann
  model for the incompressible navier--stokes equation}, Journal of Statistical
  Physics 88~(3) (1997) 927--944.
\newblock \href {https://doi.org/10.1023/B:JOSS.0000015179.12689.e4}
  {\path{doi:10.1023/B:JOSS.0000015179.12689.e4}}.
\newline\urlprefix\url{http://dx.doi.org/10.1023/B:JOSS.0000015179.12689.e4}

\bibitem{GINZBURG20051171}
I.~Ginzburg,
  \href{http://www.sciencedirect.com/science/article/pii/S0309170805000874}{Equilibrium-type
  and link-type lattice boltzmann models for generic advection and
  anisotropic-dispersion equation}, Advances in Water Resources 28~(11) (2005)
  1171 -- 1195.
\newblock \href
  {https://doi.org/https://doi.org/10.1016/j.advwatres.2005.03.004}
  {\path{doi:https://doi.org/10.1016/j.advwatres.2005.03.004}}.
\newline\urlprefix\url{http://www.sciencedirect.com/science/article/pii/S0309170805000874}

\bibitem{DHUMIERES2009823}
D.~d\'Humieres, I.~Ginzburg,
  \href{http://www.sciencedirect.com/science/article/pii/S0898122109000893}{Viscosity
  independent numerical errors for lattice boltzmann models: From recurrence
  equations to ``magic{"} collision numbers}, Computers \& Mathematics with
  Applications 58~(5) (2009) 823 -- 840, mesoscopic Methods in Engineering and
  Science.
\newblock \href {https://doi.org/https://doi.org/10.1016/j.camwa.2009.02.008}
  {\path{doi:https://doi.org/10.1016/j.camwa.2009.02.008}}.
\newline\urlprefix\url{http://www.sciencedirect.com/science/article/pii/S0898122109000893}

\bibitem{LeeT2003JCP}
T.~Lee, C.-L. Lin,
  \href{http://www.sciencedirect.com/science/article/pii/S0021999102000657}{An
  eulerian description of the streaming process in the lattice boltzmann
  equation}, Journal of Computational Physics 185~(2) (2003) 445 -- 471.
\newblock \href
  {https://doi.org/http://dx.doi.org/10.1016/S0021-9991(02)00065-7}
  {\path{doi:http://dx.doi.org/10.1016/S0021-9991(02)00065-7}}.
\newline\urlprefix\url{http://www.sciencedirect.com/science/article/pii/S0021999102000657}

\bibitem{Seta.PhysRevE89}
T.~Seta, R.~Rojas, K.~Hayashi, A.~Tomiyama,
  \href{https://link.aps.org/doi/10.1103/PhysRevE.89.023307}{Implicit-correction-based
  immersed boundary-lattice boltzmann method with two relaxation times},
  Physical Review E Statistical Nonlinear \& Soft Matter Physics 89 (2014)
  023307.
\newblock \href {https://doi.org/10.1103/PhysRevE.89.023307}
  {\path{doi:10.1103/PhysRevE.89.023307}}.
\newline\urlprefix\url{https://link.aps.org/doi/10.1103/PhysRevE.89.023307}

\bibitem{Guo2002-LBM-force}
Z.~Guo, C.~Zheng, B.~Shi, Discrete lattice effects on the forcing term in the
  lattice boltzmann method., Physical Review E Statistical Nonlinear \& Soft
  Matter Physics 65 (2002) 046308.

\bibitem{Angot1999}
P.~Angot, C.-H. Bruneau, P.~Fabrie,
  \href{http://dx.doi.org/10.1007/s002110050401}{A penalization method to take
  into account obstacles in incompressible viscous flows}, Numerische
  Mathematik 81~(4) (1999) 497--520.
\newblock \href {https://doi.org/10.1007/s002110050401}
  {\path{doi:10.1007/s002110050401}}.
\newline\urlprefix\url{http://dx.doi.org/10.1007/s002110050401}

\bibitem{Carbou2003Boundary}
Carbou, Gilles, Fabrie, Pierre, Boundary layer for a penalization method for
  viscous incompressible flow, Advances in Differential Equations 8~(12) (2003)
  1453--1480.

\bibitem{Engels201596}
T.~Engels, D.~Kolomenskiy, K.~Schneider, J.~Sesterhenn,
  \href{http://www.sciencedirect.com/science/article/pii/S0021999114006792}{Numerical
  simulation of fluid--structure interaction with the volume penalization
  method}, Journal of Computational Physics 281 (2015) 96 -- 115.
\newblock \href {https://doi.org/http://dx.doi.org/10.1016/j.jcp.2014.10.005}
  {\path{doi:http://dx.doi.org/10.1016/j.jcp.2014.10.005}}.
\newline\urlprefix\url{http://www.sciencedirect.com/science/article/pii/S0021999114006792}

\bibitem{Taylor-cylindrical-1923}
G.~I. Taylor, \href{http://www.jstor.org/stable/91148}{Stability of a viscous
  liquid contained between two rotating cylinders}, Philosophical Transactions
  of the Royal Society of London. Series A, Containing Papers of a Mathematical
  or Physical Character 223 (1923) 289--343.
\newline\urlprefix\url{http://www.jstor.org/stable/91148}

\bibitem{Niu-2006}
X.~Niu, C.~Shu, Y.~Chew, Y.~Peng,
  \href{http://www.sciencedirect.com/science/article/pii/S0375960106001472}{A
  momentum exchange-based immersed boundary-lattice boltzmann method for
  simulating incompressible viscous flows}, Physics Letters A 354~(3) (2006)
  173 -- 182.
\newblock \href
  {https://doi.org/http://dx.doi.org/10.1016/j.physleta.2006.01.060}
  {\path{doi:http://dx.doi.org/10.1016/j.physleta.2006.01.060}}.
\newline\urlprefix\url{http://www.sciencedirect.com/science/article/pii/S0375960106001472}

\bibitem{CUI201724}
X.~Cui, X.~Yao, Z.~Wang, M.~Liu,
  \href{http://www.sciencedirect.com/science/article/pii/S0021999116306726}{A
  hybrid wavelet-based adaptive immersed boundary finite-difference lattice
  boltzmann method for two-dimensional fluid--structure interaction}, Journal
  of Computational Physics 333~(Supplement C) (2017) 24 -- 48.
\newblock \href {https://doi.org/https://doi.org/10.1016/j.jcp.2016.12.019}
  {\path{doi:https://doi.org/10.1016/j.jcp.2016.12.019}}.
\newline\urlprefix\url{http://www.sciencedirect.com/science/article/pii/S0021999116306726}

\bibitem{Benson-1989}
M.~{Benson}, P.~{Bellamyknights}, J.~{Gerrard}, I.~{Gladwell}, {A viscous
  splitting algorithm applied to low Reynolds number flows round a circular
  cylinder}, Journal of Fluids and Structures 3 (1989) 439--479.
\newblock \href {https://doi.org/10.1016/S0889-9746(89)80026-X}
  {\path{doi:10.1016/S0889-9746(89)80026-X}}.

\bibitem{Ding-Shu-2004}
H.~Ding, C.~Shu, K.~Yeo, D.~Xu,
  \href{http://www.sciencedirect.com/science/article/pii/S0045782503005838}{Simulation
  of incompressible viscous flows past a circular cylinder by hybrid fd scheme
  and meshless least square-based finite difference method}, Computer Methods
  in Applied Mechanics and Engineering 193~(9 - 11) (2004) 727 -- 744.
\newblock \href {https://doi.org/http://dx.doi.org/10.1016/j.cma.2003.11.002}
  {\path{doi:http://dx.doi.org/10.1016/j.cma.2003.11.002}}.
\newline\urlprefix\url{http://www.sciencedirect.com/science/article/pii/S0045782503005838}

\bibitem{Zdravkovich-1977}
M.~M.~Zdravkovich, Review-review of flow interference between two circular
  cylinders in various arrangements, ASME Transactions Journal of Fluids
  Engineering 99 (1977) 618--633.

\bibitem{Igarashi-1981}
T.~{Igarashi}, {Characteristics of the Flow Around Two Circular Cylinders
  Arranged in Tandem: 1st Report}, JSME International Journal Series B 24
  (1981) 323--331.

\bibitem{MENEGHINI2001327}
J.~Meneghini, F.~Ssltara, C.~Siqueira, J.~Ferrari,
  \href{http://www.sciencedirect.com/science/article/pii/S0889974600903438}{Numerical
  simulation of flow interference between two circular cylinders in tandem and
  side-by-side arrangements}, Journal of Fluids and Structures 15~(2) (2001)
  327 -- 350.
\newblock \href {https://doi.org/https://doi.org/10.1006/jfls.2000.0343}
  {\path{doi:https://doi.org/10.1006/jfls.2000.0343}}.
\newline\urlprefix\url{http://www.sciencedirect.com/science/article/pii/S0889974600903438}

\bibitem{HU2014140}
Y.~Hu, H.~Yuan, S.~Shu, X.~Niu, M.~Li,
  \href{http://www.sciencedirect.com/science/article/pii/S0898122114002065}{An
  improved momentum exchanged-based immersed boundary attice boltzmann method
  by using an iterative technique}, Computers \& Mathematics with Applications
  68~(3) (2014) 140 -- 155.
\newblock \href {https://doi.org/https://doi.org/10.1016/j.camwa.2014.05.013}
  {\path{doi:https://doi.org/10.1016/j.camwa.2014.05.013}}.
\newline\urlprefix\url{http://www.sciencedirect.com/science/article/pii/S0898122114002065}

\bibitem{Wang2015}
Y.~Wang, C.~Shu, C.~Teo, J.~Wu,
  \href{http://www.sciencedirect.com/science/article/pii/S0889974614002709}{An
  immersed boundary-lattice boltzmann flux solver and its applications to
  fluid--structure interaction problems}, Journal of Fluids and Structures 54
  (2015) 440 -- 465.
\newblock \href
  {https://doi.org/http://dx.doi.org/10.1016/j.jfluidstructs.2014.12.003}
  {\path{doi:http://dx.doi.org/10.1016/j.jfluidstructs.2014.12.003}}.
\newline\urlprefix\url{http://www.sciencedirect.com/science/article/pii/S0889974614002709}

\bibitem{chan2010suppression}
A.~S. Chan, A.~Jameson,
  \href{http:https://doi.org/10.1002/fld.2075}{Suppression of the unsteady
  vortex wakes of a circular cylinder pair by a doublet-like counter-rotation},
  International Journal for Numerical Methods in Fluids 63~(1) (2010) 22--39.
\newblock \href {https://doi.org/10.1002/fld.2075}
  {\path{doi:10.1002/fld.2075}}.
\newline\urlprefix\url{http:https://doi.org/10.1002/fld.2075}

\bibitem{Chan2011Vortex}
A.~S. Chan, P.~A. Dewey, A.~Jameson, C.~Liang, A.~J. Smits, Vortex suppression
  and drag reduction in the wake of counter-rotating cylinders, Journal of
  Fluid Mechanics 679~(7) (2011) 343--382.

\bibitem{Lockard2002}
D.~P. Lockard, L.-S. Luo, S.~D. Milder, B.~A. Singer,
  \href{https://doi.org/10.1023/A:1014539411062}{Evaluation of powerflow for
  aerodynamic applications}, Journal of Statistical Physics 107~(1) (2002)
  423--478.
\newblock \href {https://doi.org/10.1023/A:1014539411062}
  {\path{doi:10.1023/A:1014539411062}}.
\newline\urlprefix\url{https://doi.org/10.1023/A:1014539411062}

\bibitem{FALAGKARIS20172348}
E.~Falagkaris, D.~Ingram, I.~Viola, K.~Markakis,
  \href{http://www.sciencedirect.com/science/article/pii/S0898122117304339}{Proteus:
  A coupled iterative force-correction immersed-boundary multi-domain cascaded
  lattice boltzmann solver}, Computers \& Mathematics with Applications 74~(10)
  (2017) 2348 -- 2368.
\newblock \href {https://doi.org/https://doi.org/10.1016/j.camwa.2017.07.016}
  {\path{doi:https://doi.org/10.1016/j.camwa.2017.07.016}}.
\newline\urlprefix\url{http://www.sciencedirect.com/science/article/pii/S0898122117304339}

\bibitem{MITTAL1994253}
S.~Mittal, T.~Tezduyar,
  \href{http://www.sciencedirect.com/science/article/pii/0045782594900299}{Massively
  parallel finite element computation of incompressible flows involving
  fluid-body interactions}, Computer Methods in Applied Mechanics and
  Engineering 112~(1) (1994) 253 -- 282.
\newblock \href {https://doi.org/https://doi.org/10.1016/0045-7825(94)90029-9}
  {\path{doi:https://doi.org/10.1016/0045-7825(94)90029-9}}.
\newline\urlprefix\url{http://www.sciencedirect.com/science/article/pii/0045782594900299}

\end{thebibliography}
\end{document}